\documentclass[lettersize,journal]{IEEEtran}
\usepackage{amsmath,amsfonts}
\usepackage{algorithmic}
\usepackage{algorithm}
\usepackage{array}
\usepackage[caption=false,font=normalsize,labelfont=sf,textfont=sf]{subfig}
\usepackage{textcomp}
\usepackage{stfloats}
\usepackage{url}
\usepackage{verbatim}
\usepackage{listings}
\lstdefinestyle{diff}{%
  basicstyle=\ttfamily\footnotesize,%
  columns=fullflexible,%
  breaklines=true,%
  keepspaces=true,%
  showstringspaces=false,%
  morecomment=[f][\color{red!70!black}]{-},%
  morecomment=[f][\color{ForestGreen}]{+}%
}
\usepackage{cite}
\usepackage{multirow}
\usepackage{makecell}
\usepackage[most]{tcolorbox}

\tcbset{
  myrqstyle/.style={
    colback=gray!5,           
    colframe=black!60,       
    arc=3pt,                 
    boxrule=1pt,           
    left=3pt,right=3pt,      
    top=2pt,bottom=2pt,      
  }
}

\usepackage[dvipsnames]{xcolor}
\newcommand{\al}[1]{\textcolor{black}{#1}}
\newcommand{\altwo}[1]{\textcolor{black}{#1}} 
\newcommand{\althree}[1]{\textcolor{black}{#1}} 

\usepackage[normalem]{ulem}
\usepackage{graphicx} 
\usepackage{amssymb}
\usepackage{booktabs}

\usepackage[table]{xcolor}
\usepackage{colortbl}
\usepackage{tikz} 

\DeclareRobustCommand{\heatcell}[3]{%
  \pgfmathsetmacro{\t}{(#3==#2) ? 0 : ((#1-#2)/(#3-#2))}%
  \pgfmathsetmacro{\t}{max(0,min(1,\t))}%
  \pgfmathsetmacro{\rval}{1 - (1-0.3882352941)*\t}%
  \pgfmathsetmacro{\gval}{1 - (1-0.7450980392)*\t}%
  \pgfmathsetmacro{\bval}{1 - (1-0.4823529412)*\t}%
  \xdef\HeatRGB{\rval,\gval,\bval}%
  \cellcolor[rgb]{\HeatRGB}#1%
}

\DeclareRobustCommand{\heatcellbf}[3]{%
  \pgfmathsetmacro{\t}{(#3==#2) ? 0 : ((#1-#2)/(#3-#2))}%
  \pgfmathsetmacro{\t}{max(0,min(1,\t))}%
  \pgfmathsetmacro{\rval}{1 - (1-0.3882352941)*\t}%
  \pgfmathsetmacro{\gval}{1 - (1-0.7450980392)*\t}%
  \pgfmathsetmacro{\bval}{1 - (1-0.4823529412)*\t}%
  \xdef\HeatRGB{\rval,\gval,\bval}%
  \cellcolor[rgb]{\HeatRGB}\textbf{#1}%
}

\begin{document}

\title{AVIATOR: Toward Agentic AI Vulnerability Injection Workflow for High-Fidelity, Large-Scale Code Security Dataset}

\author{
\textbf{Amine Lbath}$^{*\ddagger}$ \hspace{7ex} \textbf{Massih-Reza Amini}$^{\ddagger}$ \hspace{7ex} \textbf{Aurelien Delaitre}$^{*}$ \hspace{7ex} \textbf{Vadim Okun}$^{*}$
\\
\begin{minipage}[t]{0.35\textwidth}
\begin{center}
$^{*}$\textit{National Institute of Standards and Technology, Gaithersburg, MD, USA}
\end{center}
\end{minipage}
\hspace{1ex}
\begin{minipage}[t]{0.35\textwidth}
\begin{center}
$^{\ddagger}$\textit{Université Grenoble Alpes, CNRS, LIG, Grenoble, France}
\end{center}
\end{minipage}
}


\maketitle

\begin{abstract}
The increasing complexity of software systems and the sophistication of cyber-attacks have underscored the critical need for reliable automated software vulnerability detection. Data-driven approaches using deep learning models show promise but critically depend on the availability of large, accurately labeled datasets. Yet existing datasets either suffer from noisy labels, limited vulnerability coverage, or fail to reflect vulnerabilities as they occur in real-world software. This also limits large-scale benchmarking of such solutions. Automated vulnerability injection provides a way to address these limitations, but existing techniques remain limited in coverage, contextual fidelity, or injection success.
In this paper, we present AVIATOR, the first agentic AI vulnerability injection framework. AVIATOR decomposes vulnerability injection into a coordinated workflow of specialized AI agents, tool-based analysis, and iterative self-correction, explicitly mirroring expert reasoning. It integrates retrieval-augmented generation and lightweight LoRA-based fine-tuning to produce realistic, category-specific vulnerabilities without relying on handcrafted patterns.
\al{AVIATOR targets the overall quality of the generated dataset, which we assess along two main, complementary dimensions: injection fidelity and downstream detection utility.} Across three benchmarks, AVIATOR achieves \althree{a high function-level injection success rate} (91-95\%). When used for data augmentation to train deep learning-based vulnerability detection (DLVD) models, AVIATOR provides the strongest \althree{average} downstream gains in vulnerability detection\althree{, achieving the best $F_1$ score in 6 of 8 model--dataset configurations}. Across models and base datasets, AVIATOR improves average $F_1$ scores by +22\% over no augmentation, +25\% over VGX, holding the prior best injection success rate, and +3\% over VulScribeR, the prior state-of-the-art LLM-based injection model, with +7\% higher recall and no precision loss. Its augmented data exhibits the lowest distributional distortion and scales efficiently with $<$2\% syntax rejection at 4.3$\times$ lower cost than VulScribeR.
\end{abstract}


\section{Introduction}
\IEEEPARstart{T}{he} rapid growth in software complexity, coupled with the sophistication of cyber-attacks, poses a significant threat to the global security and stability of digital infrastructures. In 2024 alone, the total number of publicly reported vulnerabilities rose by 25\% \cite{cyberThreatIndex_2024}. Software vulnerabilities refer to weaknesses in system security requirements, design, implementation, or operation that could be accidentally triggered or intentionally exploited, resulting in a violation of the system's security policy \cite{sate6_delaitre2023}. These trends highlight the urgent need for scalable and effective automated vulnerability detection and repair systems. Traditional methods, such as static program analysis, have played an important role. However, their reliance on manually crafted heuristics and the combinatorial complexity of their analysis make them fundamentally limited, with persistent challenges in scalability, adaptability, and false alarm rates \cite{Not_using_Static_tools, CastleBenchmarking_dubniczky2025}.

In response, AI-based approaches for code vulnerability detection \cite{DLvulnDetectStudy_chakraborty2020, llmVulnSurvey_zhou2024, VulDeePecker_Li2018, SySeVR_Li2022, Rahman_2024, DiverseVul_yizheng2023, PrimeVul_ding2025} and repair \cite{deepcodeaifixfixing_berabi2024, vulnfixvisiontransformer_fu2025, codesecurityvulnerabilityrepair_islam2024} have recently attracted considerable attention in the research community. These data-driven techniques show strong potential in identifying and addressing a wide range of vulnerabilities. However, their effectiveness heavily relies on the quality and scale of the training data \cite{PrimeVul_ding2025, vgx_nong2024}. This dependency has become even more critical with the emergence of Large Language Models (LLMs) with billions of trainable parameters. 
For vulnerability detection in particular, the practical objective is not only to curate accurate labels, but to generate data that measurably improves downstream deep learning-based vulnerability detection (DLVD) under realistic test distributions. Yet high-quality vulnerable samples remain scarce \cite{PrimeVul_ding2025, llmVulnSurvey_zhou2024, DLvulnDetectStudy_chakraborty2020}, and existing datasets often trade off realism, label fidelity, and coverage, limiting downstream generalization.

\al{Existing vulnerability datasets range from manually and heuristically labeled real-world data to fully synthetic suites, none of which simultaneously achieves high label accuracy, scale, credibility, and diversity, as we detail in Section~\ref{subsec:related_work_datasets}. Vulnerability injection, the process of introducing realistic weaknesses into existing software~\cite{dolan2016lava, pewny2016evilcoder, vulgen_nong2023, vgx_nong2024, VulScribeR_2025}, offers a sustainable path toward this goal: it can scale to real-world codebases without manual curation and, when the injection quality is sufficient, produces accurate labels and paired benign/vulnerable samples by construction. However, existing techniques remain limited in vulnerability coverage, contextual fidelity, or injection success.}

To address these challenges, we introduce a novel agentic AI workflow designed to automatically inject realistic, diverse, category-specific vulnerabilities for large-scale generation of high-quality vulnerability datasets. This framework, called AVIATOR (agentic AI Vulnerability Injection And Transformation with Optimized Reasoning) coordinates multiple AI agents, simulating expert reasoning, alongside functional agents and traditional code analysis tools. It was designed by studying the full process of manual injection and splitting it into elementary subtasks, relying on tools and self-correction loops to provide optimal guidance to LLMs. Its design was validated by three cybersecurity experts (including two co-authors). To further boost performance, it leverages Retrieval-Augmented Generation (RAG) \cite{rag_lewis2021} for contextual grounding and employs Low-Rank Adaptation (LoRA) \cite{lora_hu2021} for efficient model fine-tuning. Crucially, AVIATOR requires no pattern mining, freeing it from the rigid constraints of mined edits and allowing scalable generation of vulnerabilities that span more Common Weakness Enumeration (CWE) categories, with greater semantic and contextual realism.

\al{We assess the quality of the datasets generated by AVIATOR along two complementary dimensions: injection success rate, which reflects whether the generated samples are correctly labeled, and downstream utility, which reflects whether the generated data is realistic and diverse enough for detection models to generalize to real-world vulnerabilities. Prior generation methods pursue the same goal of high-quality data but tend to foreground a single dimension, VGX~\cite{vgx_nong2024} injection success and VulScribeR~\cite{VulScribeR_2025} downstream detection, even though neither alone fully reflects dataset quality. AVIATOR instead targets overall sample quality directly, producing samples that are correctly instantiated, realistic, and diverse, and it improves on the strongest prior method along both dimensions, indicating higher-quality data across the board.}

\al{To evaluate the first dimension,} we conducted an experimental study on code samples from three widely used benchmarks \cite{nist_sard_2018, formai_tihanyi2023, PrimeVul_ding2025}. We used the formal verification tool ESBMC \cite{esbmc_gadelha2018} to validate supported vulnerabilities in smaller, self-contained functions, and we relied on manual evaluation by experienced security researchers for larger functions with missing project-level context, where ESBMC is not applicable. The results show that AVIATOR achieves highest injection fidelity, with success rates between 91\% and 95\% across the evaluated benchmarks\al{. We use injection fidelity to denote label correctness, namely whether a generated sample actually instantiates the intended weakness \althree{ within the function body}, which inherently corresponds to the label accuracy of the resulting dataset}.

Beyond injection fidelity, AVIATOR also delivers the strongest downstream gains under the same DLVD augmentation protocol as VulScribeR \cite{VulScribeR_2025}\al{, surpassing this prior state of the art in average $F_1$ score \althree{(best $F_1$ in 6 of 8 model--dataset configurations)} while increasing recall by \textbf{+7.4\%} with no loss in precision. This profile is} consistent with cleaner injected vulnerabilities and a more learnable training signal\al{: detectors recover substantially more true vulnerabilities without raising false alarms}. Among all augmentation strategies, AVIATOR also induces the smallest distributional shift from the original real-world dataset, indicating that it enriches training data without skewing it toward narrow templates. \al{Together with its high injection fidelity, this makes }AVIATOR suitable not only for augmentation but also for constructing benchmark datasets with reliable labels.

AVIATOR is a generic framework capable of supporting injection across a large CWE spectrum. Unlike prior injection-based approaches, AVIATOR does not rely on handcrafted pattern mining. Instead, its agentic design leverages LLM-based reasoning, making the workflow broadly applicable and easier to adapt across programming languages. \al{It also operates efficiently at scale, with a low syntax rejection rate and a generation cost several times lower than comparable LLM-based pipelines,} enabling practical large-scale dataset construction.

The key contributions of this work are:
\begin{itemize}
    \item \textbf{AVIATOR: Agentic Vulnerability Injection Framework.}
    To the best of our knowledge, this is the first work to explore vulnerability injection using an agentic AI workflow. We carefully designed a multi-agent pipeline, leveraging tool-augmented subtasks with self-correction and validation loops, that replicates the reasoning processes of cybersecurity experts.

    \item \textbf{High-Fidelity, Broad-Coverage Vulnerability Injection.}
    Our evaluation demonstrates that AVIATOR achieves between 91\% and 95\% success in generating plausible vulnerabilities across three diverse benchmarks. AVIATOR is designed as a generic framework capable of supporting injection across a wide spectrum of weakness categories; our current instantiation supports \textbf{133 CWE types}, far exceeding state-of-the-art systems like VGX.  This demonstrates a strong potential for constructing diverse, high-coverage datasets.

    \item \textbf{Downstream Effectiveness.}
    AVIATOR improves the average $F_1$-score \al{for vulnerability detection} by \textbf{+21.8\%} over no augmentation and \textbf{+24.6\%} over VGX. Notably, it surpasses the LLM-based state-of-the-art \textbf{VulScribeR} by \textbf{+2.9\%} in average $F_1$ score and \textbf{+7.4\%} in recall, achieving these gains without degrading precision.

    \item \textbf{Realistic Augmentation and Practical Scalability.}
    We show that AVIATOR induces the smallest distributional distortion relative to the original training set (lowest KL divergence) among augmentation strategies, indicating more realistic augmentation. AVIATOR also scales efficiently, maintaining a $\mathbf{<2\%}$ syntax rejection rate and achieving a \textbf{4.3$\times$} lower generation cost than the LLM-based technique VulScribeR.
\end{itemize}

The remainder of this paper is organized as follows: \al{Section~2 surveys the background and related work on vulnerability datasets, detection models, and injection techniques.} Section~3 details the architecture and agentic workflow of AVIATOR. Section~4 outlines our experimental and evaluation setup. Section~5 reports empirical results and analysis. Section~6 discusses limitations and future work, and Section~7 concludes.

\section{\al{Background and Related Work}}
\label{sec:background}
\al{This section reviews the background concepts and related literature that motivate and situate AVIATOR. We begin with deep learning-based vulnerability detection (Section~\ref{subsec:related_work_dlvd}), which drives our work yet remains limited by data scarcity. We then define the properties of high-quality vulnerability datasets (Section~\ref{subsec:properties_high_qual_vuln_dataset}) and survey existing datasets (Section~\ref{subsec:related_work_datasets}) and automated injection and augmentation techniques (Section~\ref{subsec:related_work_injection}). We conclude with a formal definition of the vulnerability injection problem (Section~\ref{subsec:injection}).}

\subsection{Deep Learning-Based Vulnerability Detection}
\label{subsec:related_work_dlvd}

\al{Deep learning has emerged as a promising solution for automated vulnerability detection, with approaches differing primarily in how they represent code and the neural architecture employed.}

\al{\textbf{Sequence-based models} were among the first deep learning approaches to vulnerability detection. VulDeePecker~\cite{VulDeePecker_Li2018} and SySeVR~\cite{SySeVR_Li2022} extract program slices ("code gadgets") around data and control dependencies and classify them with bidirectional recurrent networks. These methods established the viability of learning vulnerability patterns from code but are sensitive to how slices are constructed and offer limited contextual reasoning.}

\al{\textbf{Graph-based models} represent code as graphs capturing control- and data-flow. Devign~\cite{devign_zhou2019} applies a gated graph neural network to a joint graph of abstract syntax tree, control-flow, and data-flow, while ReVeal~\cite{DLvulnDetectStudy_chakraborty2020} uses Code Property Graphs with representation learning and resampling to counter class imbalance, and FGVulDet~\cite{fgvuldet_liu2024} adds edge-type awareness for fine-grained, type-specific detection. Such models better capture structural patterns but still fail to generalize well enough \cite{DiverseVul_yizheng2023, PrimeVul_ding2025}.}

\al{\textbf{Pre-trained transformer models} leverage large-scale code pretraining: VulBERTa~\cite{vulberta_hanif2022} uses a RoBERTa model with a code-specific tokenizer, LineVul~\cite{linevul_fu2022} fine-tunes a BERT-based model for line-level prediction, and CodeT5~\cite{codet5_wang2021} is an encoder-decoder model widely used as a detection baseline~\cite{PrimeVul_ding2025}; Kalouptsoglou et al.~\cite{transferlearning_kalouptsoglou2025} find that extracting features from such models can match full fine-tuning at far lower training cost. These models, however, tend to rely on superficial token patterns rather than vulnerability semantics, generalize poorly out of distribution, and are limited by a fixed context window that truncates long functions~\cite{Rahman_2024, PrimeVul_ding2025}.}

\al{\textbf{Large language models} represent the current frontier~\cite{llmVulnSurvey_zhou2024, PrimeVul_ding2025}, with recent work also targeting robustness, e.g., causal learning to reduce reliance on spurious correlations~\cite{Rahman_2024}.}

\al{While earlier methods were often limited by the expressiveness of their own representations, the rapid growth in model capacity has shifted the binding constraint: across the recent literature, the emerging consensus is that further progress is now limited primarily by the scarcity of high-quality, accurately labeled training data rather than by model capacity~\cite{llmVulnSurvey_zhou2024, PrimeVul_ding2025, DataQuality_croft2023, dataprep_croft2023, DLvulnDetectStudy_chakraborty2020}. This data bottleneck motivates a closer look at what constitutes a high-quality vulnerability dataset, and at how such datasets can be constructed at scale.}

\subsection{Properties of High-Quality Vulnerability Datasets}
\label{subsec:properties_high_qual_vuln_dataset}

Effective training and rigorous benchmarking of DLVD methods critically depend on the quality of the underlying datasets. To clearly define the desired dataset characteristics, we introduce the ABCD criteria, similar to \cite{sate6_delaitre2023} and \cite{dolan2016lava}:

\begin{itemize}
\item \textbf{Accurate}: Labels must reflect ground truth with clear information on the weakness type (CWE identifiers) and, when possible, precise weakness locations, to support more reliable training and faithful evaluation.

\item \textbf{Big}: The dataset should be large enough to support training of deep learning models and enable statistically meaningful benchmarking of detection and repair systems.

\item \textbf{Credible}: Code samples should reflect the structure and complexity of real-world software to ensure model generalization in practical settings.

\item \textbf{Diverse}: The dataset should cover varied categories of weaknesses and coding styles to enable generalizable learning and comprehensive benchmarking across scenarios.
\end{itemize}

In practice, additional constraints improve training utility:
\begin{itemize}
\item Each sample should ideally contain a single weakness type to reduce ambiguity during training and evaluation.
\item Paired benign and vulnerable samples are valuable for repair tasks and contrastive learning.
\end{itemize}

\subsection{\al{Vulnerability Datasets}}
\label{subsec:related_work_datasets}

\al{Constructing vulnerability datasets that satisfy all four ABCD properties remains a persistent open challenge~\cite{llmVulnSurvey_zhou2024, DataQuality_croft2023, dataprep_croft2023}. We review the main curation-based strategies below, each with characteristic trade-offs, before turning to injection-based generation in Section~\ref{subsec:related_work_injection}.}

\textbf{Synthetic datasets}, such as the Juliet Test Suite~\cite{juliet2017} \althree{ and, more recently, LLM-synthesized samples~\cite{titanvul_li2026}}, provide clean, controllable examples with more accurate labels and broad CWE coverage. However, they rarely reflect the structural and semantic complexity of production code. Models trained on such data often learn superficial cues rather than underlying vulnerability patterns~\cite{DLvulnDetectStudy_chakraborty2020}.

\textbf{Heuristic-labeled real-world datasets} such as ReVeal~\cite{DLvulnDetectStudy_chakraborty2020}, BigVul~\cite{bigvul_fan2020}, CVEfixes~\cite{CVEfixes_guru2021}, CrossVul~\cite{CrossVul_georgios2021}, and DiverseVul~\cite{DiverseVul_yizheng2023} are mined from vulnerability-fixing commits and annotated using security databases such as NVD~\cite{nist_nvd_2022}. While they capture real-world code complexity and diversity, their heuristic labeling introduces substantial noise, with reported label accuracies as low as 25\% for some datasets \cite{PrimeVul_ding2025, DataQuality_croft2023}. Static-analyzer-labeled datasets (e.g., D2A~\cite{d2adatasetbuilt_zheng2021}, Draper~\cite{draper_russell2018}) suffer similarly from high false-positive rates due to tool limitations.

\textbf{Reliable-label datasets} \al{address the accuracy dimension through stricter controls}. Manual annotation (e.g., SVEN~\cite{SVEN_he2023}) improves label fidelity but remains expensive and small-scale. Conversely, PrimeVul~\cite{PrimeVul_ding2025} improves reliability by applying strict heuristics to vulnerability-fixing commits, reaching human-level labeling accuracy, but at the cost of scalability and coverage. Neither route offers a robust path to producing large, high-quality vulnerable data at scale.

\al{Each strategy trades off differently: synthetic generation yields accurate labels but unrealistic code, whereas curating real code preserves realism at the expense of label accuracy or scale. Vulnerability injection has the potential to combine both strengths by introducing weaknesses directly into real software, which we examine next.}

\subsection{Automated Vulnerability Injection and Data Augmentation}
\label{subsec:related_work_injection}

Vulnerability injection offers a scalable route to constructing labeled datasets by introducing realistic weaknesses into existing code bases. Early tools such as LAVA~\cite{dolan2016lava} and EvilCoder~\cite{pewny2016evilcoder}, along with the bug-synthesis approaches of Roy et al.~\cite{roy2018bug} and Kashyap et al.~\cite{kashyap2019automated}, demonstrated the feasibility of automated bug insertion but can produce synthetic-looking code that diverge from real development patterns.

\textbf{Pattern-mining approaches} combine mined vulnerability templates with deep learning-based localization to target real code. VULGEN~\cite{vulgen_nong2023} mines vulnerability-fixing commits to extract edit patterns and applies a neural model to identify suitable injection sites in new functions. VinJ~\cite{vinj_nong2024} extends this into a practical, scalable tool. Both improve code credibility over early tools but remain constrained by the expressiveness of mined patterns, reporting injection success rates of approximately 69\%.

VGX~\cite{vgx_nong2024} advances the pattern-mining paradigm with a more principled site-selection strategy and achieves higher injection success (approximately 90\%). However, it relies on a partly manual pattern-mining pipeline, focuses on single-statement edits, and supports only 23 CWE types, restricting the diversity of the generated dataset.

\al{\textbf{Vulnerability-preserving data augmentation} multiplies existing vulnerable samples rather than creating new ones. Liu et al.~\cite{fgvuldet_liu2024} apply label-preserving code mutations (e.g., identifier renaming and statement reordering) to expand scarce vulnerability types without altering their semantics. However, these edits are surface-level and operate only on already-vulnerable code, so the augmented samples stay structurally close to their seeds and cannot introduce new vulnerability patterns or broaden CWE coverage, limiting the diversity and expressiveness of the resulting data.}

\textbf{LLM-based injection} lifts these template-bound constraints by leveraging the generative capabilities of large language models. VulScribeR~\cite{VulScribeR_2025} combines an LLM with retrieval-augmented generation (RAG) to generate augmentation data and primarily reports strong downstream DLVD gains, with broader CWE coverage and more natural code than pattern-mining. However, its injection success remains more limited (an estimated 82\%) and can depend on model variability and post-filtering; the resulting label noise both weakens its suitability for benchmark construction and can cap the downstream gains achievable when training on the data.

These approaches expose two complementary but often disconnected facets of dataset quality. The first is \emph{label fidelity}, whether the generated samples are correctly labeled, as approximated by injection success. The second is \emph{realism}, whether the data is natural and diverse enough for a detector trained on it to generalize to real vulnerabilities, as approximated by downstream detection performance. Prior work tends to advance only one facet at a time: VGX emphasizes injection success under narrow, template-like edits, whereas VulScribeR emphasizes downstream detection. \al{Rather than optimizing either proxy in isolation, AVIATOR targets dataset quality itself through an agentic AI workflow that mirrors expert reasoning and couples multi-step validation with lightweight fine-tuning; both facets then improve as a consequence rather than as separately tuned objectives.}

\subsection{Vulnerability Injection}
\label{subsec:injection}

We consider a code snippet \(c\) as an ordered sequence of lexical tokens $c = \langle c_1, c_2, \dots, c_n\rangle,\quad c_i \in T$, where \(T\) is the vocabulary of the programming language (e.g., operators, delimiters, identifiers, literals, keywords, punctuation), and the token sequence conforms to the language grammar.

We define a vulnerability injection operator:
{\small
\begin{equation}
  \mathsf{inject}: T^n \times V \longrightarrow T^m,\qquad
  \bigl(c^b,v_j\bigr) \mapsto c^{\mathrm{v}}
  \label{eq:vul_inj_func}
\end{equation}
}
which transforms a benign snippet $c^b$ into a vulnerable snippet $c^v$ exhibiting the vulnerability $v_j \in V$, where $V$ is the set of existing vulnerability types.

Injection is performed through a finite set of token-level edits. Let $ I = \{\,i_1,\dots,i_k\}\subseteq\{1,\dots,n\}$ be the set of positions to modify, and let each edit be defined by a function
{\small
\[
  \tau_i: T \longrightarrow T^*, \quad \tau_i(c^b_{(i)}) = \langle t^{(i)}_1, \dots, t^{(i)}_{m^{(i)}} \rangle,
\]
}
where \(m^{(i)} \ge 0\), and \(T^*\) is the set of finite token sequences. This formulation captures:

\begin{itemize}
  \item \textbf{Insertion:} New tokens are inserted before or after \(c^b_{(i)}\):
  {\small
  \[\text{For~}m^{(i)} \geqslant 1; \begin{cases}
    \tau_i(c^b_{(i)}) = \langle t^{(i)}_1, \dots, t^{(i)}_{m^{(i)}}, c^b_{(i)} \rangle, & \text{or}\\
    
    \tau_i(c^b_{(i)}) = \langle c^b_{(i)}, t^{(i)}_1, \dots, t^{(i)}_{m^{(i)}} \rangle. & 
  \end{cases}
  \]
  }
    \item \textbf{Deletion:} \(c^b_{(i)}\) is removed entirely:
    {\small
  \[
    \quad \tau_i(c^b_{(i)}) = \langle\rangle,\quad m^{(i)} = 0.
  \]
    }
    \item \textbf{Replacement:} $c^b_{(i)}$ is replaced by a new non-empty token sequence:
    {\small
  \[
    \tau_i(c^b_{(i)}) = \langle t^{(i)}_1,\dots,t^{(i)}_{m^{(i)}} \rangle,\quad m^{(i)} \geqslant 1.
  \]
    }
\end{itemize}

For positions \(i \notin I\), we define \(\tau_i(c^b_{(i)}) = \langle c^b_{(i)} \rangle\), leaving the token unchanged.

The final injected sequence is obtained by flattening all transformed segments:
{\small
\begin{equation}
  \label{eq:vuln_inj_seq_op}
  c^v = \bigoplus_{i=1}^n \tau_i(c^b_{(i)}) = \langle c'_1, c'_2, \dots, c'_m \rangle,
\end{equation}
}
where \(\bigoplus\) denotes sequence concatenation.

High-quality injection must satisfy the following properties:
\begin{itemize}
  \item \textbf{Realism:} Edits \(\{\tau_i\}\) should reflect patterns observed in real-world vulnerabilities of class \(v_j\) (e.g. adapt to existing dataflow, meaningful output, no dead storage ...).
  \item \textbf{Functional Integrity:} $c^v$ must remain syntactically and semantically valid, preserving the original functionality of $(c^b)$, aside from the injected flaw.
  \item \textbf{Minimality:} Number of edits \(|I|\) should be kept minimal.
\end{itemize}

\section{Proposed Framework}

\begin{figure*}[!ht]
    \centering
    \includegraphics[width=0.9\textwidth]{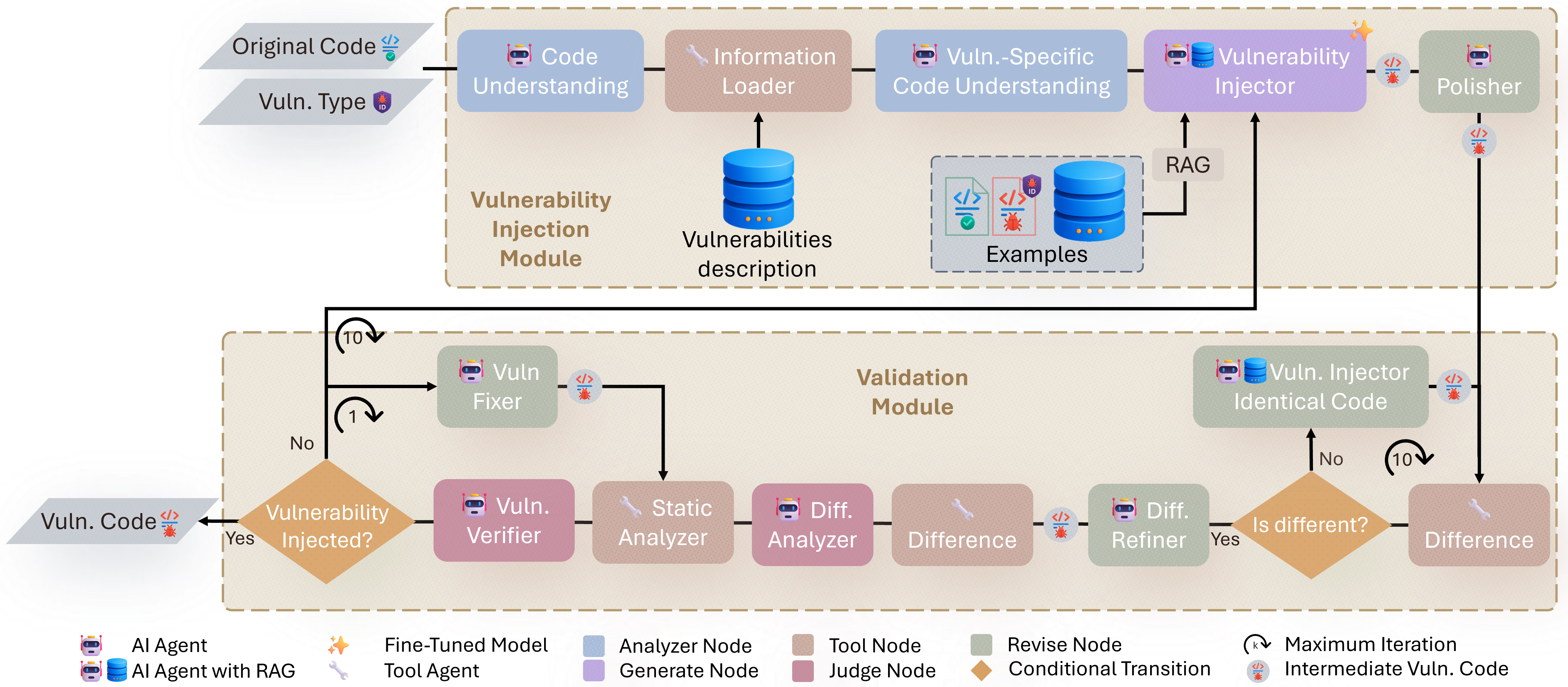}
    \caption{Architecture of the AVIATOR framework illustrating the two core modules. The top module represents the Vulnerability Injection component, which utilizes RAG and supervised fine-tuning with LoRA for guiding code transformations. The bottom module depicts the Validation component, combining LLM-based discriminators with static analysis to verify vulnerability instantiation and trigger bounded self-correction. The full framework contains 13 agents including 9 AI agents and 4 tool agents.}
    \label{fig:aviator_architecture}
\end{figure*}

Existing vulnerability injection techniques trade off fidelity, coverage, and downstream utility. Pattern-mining methods such as VulGen, VinJ, and VGX scale well but are constrained to narrow edit templates and limited CWE coverage, while recent LLM-based approaches such as VulScribeR reach broader coverage but lower injection success. To address these limitations, we propose \textbf{AVIATOR} (Figure~\ref{fig:aviator_architecture}), which formulates vulnerability injection as an \emph{agentic AI workflow} that explicitly mirrors the structured reasoning process of a human security analyst. \al{Rather than optimizing injection correctness or downstream performance on its own, it targets the overall quality of the generated dataset.}

\al{At a high level, AVIATOR takes a benign function and a target CWE, and introduces targeted changes that instantiate that weakness. Each pass thus turns the function into one labeled vulnerable sample, and repeating this across a corpus yields the new samples used to augment vulnerability-detection datasets (Section~\ref{subsec:downstream_setup}).}

This section details the design principles underlying AVIATOR, the agentic workflow abstraction on which it is built, and the two core modules of the framework: vulnerability injection and validation.

\subsection{Agentic AI Workflows}
\label{subsec:AI_agentic_workflows}

AVIATOR builds on the concept of an \emph{agentic workflow}, where a complex task is decomposed into smaller, well-defined subtasks executed by specialized agents. Each agent is instantiated as either a large language model (LLM) or a programmatic tool such as a static analyzer. Agents are connected through a directed execution graph with typed inputs and outputs, and conditional transitions that allow for retries or fallbacks when earlier steps fail. In contrast to monolithic program transformations that rely on a single model to perform injection end-to-end, this design constrains the scope of each agent and enables structured feedback between stages.

Formally, an agentic workflow $W_k$ with $k$ agents is a directed graph where nodes correspond to agents and edges represent conditional transitions. Each agent $A_i$ is defined as a transformation $A_i : X_i \times C_i \mapsto Y_i$, where $X_i$ is the structured input, $C_i$ is an optional set of contextual artifacts (e.g., CWE descriptions, retrieved examples, static analysis reports), and $Y_i$ is the output. Transitions are governed by conditions dynamically evaluated from prior outputs, allowing the workflow to re-invoke or bypass agents as needed.

The agentic workflow design provides advantages over monolithic reasoning or linear pipelines. First, it enables modular decomposition: the overall task is split into smaller, logically coherent subtasks aligned with the competencies of individual agents. This reduces the reasoning burden on any single component and improves interpretability. Second, it enables structured context propagation: each agent receives only the relevant inputs for its subtask, while outputs from earlier agents can be enriched with retrieved or validation results. Third, agentic workflows support corrective execution: if an agent detects inconsistencies or failures in prior outputs, the workflow can re-invoke earlier agents to revise decisions or recover from errors. This modular design promotes targeted decision-making, minimizes error propagation, and ensures that vulnerability injection proceeds in a systematic, verifiable, and context-aware manner. These properties make agentic workflows well suited to complex code transformation tasks such as vulnerability injection.

\subsection{Design Methodology and Iterative Refinement}
\label{subsec:design_methodology}

AVIATOR's workflow design was guided by expert knowledge and iterative refinement. We interviewed three experienced cybersecurity researchers (including two co-authors) to document how they would manually inject vulnerabilities. Their reasoning consistently followed four stages: (1) understand the program and its dataflow, (2) identify a suitable injection site (e.g. sanitization checks preventing intended flaw), (3) introduce a subtle but plausible modification, and (4) review the change for plausibility and CWE alignment. These stages formed AVIATOR's high-level structure.

Each of these stages was decomposed into elementary subtasks and mapped to specialized agents. For instance, comprehension was assigned to the \emph{Code Understanding} and \emph{Vuln.-Specific Code Understanding} agents, and review to a combination of diff-based agents and static analysis. This decomposition reduces the burden on individual agents and integrates LLMs with deterministic tools.

The workflow was then refined through repeated experimentation on the SARD test suite 100 (described Section~\ref{subsec:datasets}), which served as our primary development benchmark. Early prototypes exposed recurring issues, including truncated code, over-commented edits, and trivial modifications such as whitespace changes. To address these failure modes, we introduced new agents incrementally. Examples include the \emph{Polisher}, designed to normalize formatting and repair truncated outputs, and the diff-based refinement loop, which detects trivial edits and triggers corrective re-injection.

Crucially, these refinements targeted generic error patterns rather than dataset-specific artifacts. By addressing issues such as non-substantive edits and truncation, we ensured that improvements generalized across datasets. As shown later Section~\ref{sec:results}, the same workflow behavior transfers to distinct datasets, confirming that the iterative process captured robust error-recovery patterns rather than overfitting to SARD 100.

Through this iterative design, AVIATOR emerged as a workflow organized into two complementary modules (Figure~\ref{fig:aviator_architecture}). The \textbf{Vulnerability Injection Module} analyzes benign code, identifies potential injection points, and introduces a vulnerability. The \textbf{Validation Module} then checks these candidates using difference analysis, static analysis, and contextual review, providing structured feedback for correction or reinjection. Together, these two modules instantiate the reasoning process elicited from experts: understanding, modifying, reviewing, and revising.

\subsection{Vulnerability Injection Module}
\label{subsec:vuln_inj_module}

The Vulnerability Injection Module converts a benign function $c^b$ into a candidate vulnerable variant $c^v$ of a specified CWE. It instantiates the first three stages of the analyst-inspired workflow: comprehension, vulnerability-specific analysis, and transformation. The module comprises the agents \al{that we describe below}. \al{They operate sequentially, each building on the context produced by the earlier agents} (Figure~\ref{fig:aviator_architecture}). We provide the exact prompt templates used by the
major agents in AVIATOR in Appendix~\ref{appendix:prompts}.

\subsubsection{Code Understanding (LLM)}\label{sussubsec:code_understanding} The process begins with semantic and structural analysis of the benign code. \al{Taking the benign function $c^b$ as its sole input, the LLM} agent qualitatively analyzes \al{its} control and data flow to infer its functional intent, input/output behavior, and internal logic. It also infers potential \textit{sources}—locations where untrusted input may enter the program—and \textit{sinks}—locations where sensitive operations occur. \altwo{These descriptions are qualitative summaries that provide enhanced context to the subsequent agents. We leverage model reasoning at this step, in place of deterministic tools, because AVIATOR operates at the function level, where identifying intent, sources, and sinks depends on what the code achieves rather than on its structure alone. It also keeps this stage free of language-specific tooling, supporting the portability discussed in Section~\ref{subsec:other_languages}.} This stage mirrors the comprehension step a human analyst performs.
\subsubsection{Information Loader (tool)}\label{sussubsec:information_loader} Retrieves a detailed definition and description of the type of weakness that needs to be injected\al{, drawn from a prepared catalog of CWE descriptions (Section~\ref{subsec:implementation})}. This helps ensure the LLM receives proper context on this weakness category.
\subsubsection{Vuln.-Specific Code Understanding (LLM)}\label{sussubsec:sanitization_detector} 
Examines the benign code to identify elements that prevent the intended vulnerability from manifesting. These include explicit sanitization checks (e.g., bounds validation) as well as broader safety mechanisms such as correct memory management, error handling, or defensive programming patterns. The agent produces a structured inventory of these safeguards, which downstream agents use to determine how the code can be realistically transformed into a vulnerable variant. This process mirrors traditional symbolic static analysis and taint-tracking techniques \cite{staticAnalysis_schwartz2010}, but applied inversely. Instead of proving correctness, it highlights the logic that enforces safety. This reverse reasoning is critical to selecting feasible and non-trivial injection points.
\subsubsection{Vulnerability Injector (LLM + RAG + fine-tuning)} Relies on previous information to apply the transformation and justify its injection. To further boost the performance of this critical component we leverage retrieval-augmented generation (RAG) as well as fine-tuning. In the following, we first explain the retrieval-augmented vulnerability generation, then detail the low-rank adaptation fine-tuning method, followed by supervised fine-tuning and reinforcement learning strategies.
\paragraph*{Retrieval-Augmented Vulnerability Generation}
Once candidate injection points are identified, an agent transforms the code to inject the target weakness (\textit{Vulnerability Injector} agent). This agent operates under a retrieval-augmented generation (RAG) setup. Given an input function \( c^b \), the agent retrieves a set of $k$ similar benign/vulnerable pairs \( \{ c^{b_{(j)}}, c^{v_{(j)}} \}_{j=1}^k \) from a knowledge base \al{of aligned benign/vulnerable pairs (Section~\ref{subsec:datasets})} using embedding-based similarity. Retrieved examples are aligned using line-level diffs, marking the injected edits.

This retrieved context alongside the analyzed output, is provided to the LLM to guide code transformation. This ensures that the injection edits reflect realistic code structure and vulnerability patterns observed in real-world examples.
\paragraph*{Low-Rank Adaptation (LoRA)}
To specialize the vulnerability injection agent without incurring the computational cost of full model fine-tuning, we apply \textit{Low-Rank Adaptation (LoRA)} \cite{lora_hu2021}. This technique is particularly effective in settings where the base model is large, but the task-specific dataset is limited in size. Rather than updating the full weight matrix \( W_0 \in \mathbb{R}^{d \times k} \), LoRA learns a low-rank decomposition by injecting two trainable matrices \( B \in \mathbb{R}^{d \times r} \) and \( A \in \mathbb{R}^{r \times k} \), with \( r \ll \min(d, k) \), into the model's architecture: $W = W_0 + BA$. This reduces the number of trainable parameters while retaining the benefits of large-scale pretraining.

We explore two distinct fine-tuning paradigms for the vulnerability injection agent: supervised fine-tuning (SFT) and reinforcement learning with Group Relative Policy Optimization (GRPO). Both approaches operate on the same dataset of aligned benign/vulnerable function pairs \al{(Section~\ref{subsec:datasets})}:
\begin{equation}
\mathcal{Q} = \left\{ \left(c^{b_{(j)}},\, \hat c^{v_{(j)}} \right) \right\}_{j=1}^N
\label{eq:dataset}
\end{equation}
where each benign function $c^{b_{(j)}}$ is accompanied by a structured prompt $q^{(j)}$. The prompt is generated by using the output of prior workflow steps \ref{sussubsec:code_understanding}, \ref{sussubsec:information_loader}, \ref{sussubsec:sanitization_detector}. This prompt incorporates (1) semantic and structural analysis results of the benign code, (2) the original benign function itself, and (3) relevant examples from the Retrieval-Augmented Generation (RAG) knowledge base, annotated to highlight vulnerability injection locations. The objective is for the agent to generate the corresponding vulnerable function \( \hat c^{v}_{j} \) given this prompt.

\paragraph*{Supervised Fine-Tuning (SFT)}
The vulnerability injection agent is first trained via supervised fine-tuning (SFT), using a standard autoregressive next-token prediction objective. 
Let \(q^{(j)} \) denote the structured prompt. For example \( c^{b_{(i)}} \) and \( \hat{\mathbf{c}}^{{v}_{i}} = \langle y^{(i)}_1, \dots, y^{(i)}_{T^{(i)}} \rangle \) the tokenized target vulnerable function of size $T^{(i)}$. We fine-tune the LoRA adapter parameters \(\phi\) by minimizing the negative log-likelihood of each next token under teacher forcing. The loss for a batch of size \( B \) is:
{\small
\begin{equation}
\label{eq:sft}
\mathcal{L}_{\text{SFT}}(\phi) 
= - \frac{1}{B} \sum_{i=1}^{B} \sum_{t=0}^{T_i-1} 
\log P\bigl(y_{i,t+1}\,\bigm\vert\,\mathbf{x}_i,\,
y_{i,1},\dots,y_{i,t};\,\phi\bigr).
\end{equation}
}
Here, \(P(\cdot\mid \cdot;\phi)\) is the autoregressive distribution defined by the LoRA-adapted LLM. This technique focuses on exact token-level matching with the aim that it will learn underlying concepts behind vulnerability injection, guided explicitly by ground-truth examples.

\paragraph*{Group Relative Policy Optimization (GRPO)}
While SFT optimizes for token-level accuracy, it does not directly focus on structural and semantic nuances of code. Therefore, we also apply reinforcement learning (RL) via \textit{Group Relative Policy Optimization (GRPO)}~\cite{grpo_shao2024}, using the CodeBLEU~\cite{codebleu_ren2020} metric as a reward function. 

GRPO is an efficient reinforcement learning technique that eliminates the need for a separate critic network by comparing multiple candidate outputs for each prompt and computing their relative rewards within a sampled group. Given a prompt \( q \), we sample \( G \) candidate outputs \( \{o_1, o_2, \dots, o_G\} \) from the old policy model \( \pi_{\theta_{\text{old}}} \). Each candidate output \( o_i \) receives a reward score \( r(o_i) \) based on the CodeBLEU metric, which better captures structural and semantic properties of code. CodeBLEU combines four components:
{\small
\begin{itemize}
  \item \textbf{N-gram match}: standard BLEU score over tokens.
  \item \textbf{Weighted n-gram match}: gives higher importance to syntax-critical tokens.
  \item \textbf{AST match}: measures similarity of abstract syntax trees.
  \item \textbf{Data-flow match}: compares data-flow graphs to assess semantic equivalence.
\end{itemize}
\begin{multline}
\label{eq:codebleu}
\text{CodeBLEU} = \alpha \cdot \text{BLEU}(n\text{-grams}) 
+ \beta \cdot \text{BLEU}_{w}(n\text{-grams}) \\
+ \gamma \cdot \text{Match}_{\text{AST}} 
+ \delta \cdot \text{Match}_{\text{Data-flow}},  
\end{multline}
}
with $\alpha + \beta + \gamma + \delta = 1$. By default, $\alpha = \beta = \gamma = \delta$ \cite{codebleu_ren2020}.

\al{Following the standard GRPO formulation~\cite{grpo_shao2024}, each candidate's CodeBLEU reward is normalized into a group-relative advantage, and the policy is updated with a clipped surrogate objective regularized by a KL penalty toward the pretrained reference model; we give the full objective in Appendix~\ref{appendix:grpo}. This lets us test whether a structure-aware reward improves injection beyond token-level SFT.}
\subsubsection{Polisher (LLM)} After a first version of the injection is completed, this agent was introduced to fix common issues from the \textit{Vulnerability Injector}. It helps to normalize formatting, corrects truncated outputs, and removes superficial edits such as comment-only changes. The polished candidate is then handed to the Validation Module.

\subsection{Validation Module}
\label{subsec:valid_module}

The Validation Module \al{receives the candidate $c^v$ produced by the Injection Module} and ensures that it meaningfully introduces the intended vulnerability into the original function $c^b$. It mirrors the review stage of human analysts: checking plausibility and revising when necessary. The module combines difference analysis, contextual LLM judgment, static analysis, and bounded correction loops. \al{Similarly to the previous module, each agent draws on the context generated by prior agents and follows the flow depicted in Figure~\ref{fig:aviator_architecture}}.

Given $(c^b, c^v)$, compute a line diff. If there are no changes or the changes are only whitespace/comments, the injection failed. In that case, the workflow is routed to the \emph{Vuln.\ Injector Identical Code}. Otherwise, if there is at least one non-trivial edit, the candidate proceeds directly to the \emph{Diff Refiner}.
\subsubsection{Vuln.\ Injector Identical Code (LLM)}
Invoked only when the triviality gate detects no or trivial edits only. This agent re-attempts to generate $c^v$ from scratch, incorporating the context provided by prior agents, with the aim of increasing the likelihood that the agent actually introduces the weakness this time. The new candidate is returned to the \emph{Difference (Triviality Gate)} for re-evaluation. This loop is bounded (default: 10 attempts) to prevent infinite cycling.
\subsubsection{Diff Refiner (LLM)}
For candidates with substantive edits, the Diff Refiner analyzes the generated code together with the annotated difference and determines whether the changes correctly implement the requested weakness. 
If the edits are incomplete, incorrect, or misaligned, the agent updates the vulnerable code accordingly. 
Its judgment draws on the context accumulated from all previous agents, ensuring that the revised $c^v$ both preserves original functionality and aligns with the CWE intent before proceeding to downstream validation.
\subsubsection{Difference (tool): Reviewer Diff}
The revised $c^v$ is diffed again to produce the reviewer-oriented, annotated difference consumed by following analyzers.
\subsubsection{Diff Analyzer (LLM)}
The Diff Analyzer serves as a second reviewer to double-check the edits introduced in $c^v$. 
It analyzes the generated code together with the annotated difference and reassesses whether the modifications correctly introduce the requested weakness while preserving the original functionality.
\subsubsection{Lightweight Static Analyzer (tool)}
Runs lightweight static analysis on $c^v$ using \texttt{Cppcheck}\footnote{\url{https://cppcheck.sourceforge.io/}}. Diagnostics are passed to the verifier. While we use \texttt{Cppcheck} for C/C++, this component is generic and can be swapped with a static analyzer appropriate for other languages.
\subsubsection{Vulnerability Verifier (LLM)}
Integrates diagnostics from the static analyzer with the annotated diff and CWE context to decide whether the candidate indeed reflects the intended flaw. 
On success, the module outputs the validated $c^v$; otherwise, control transfers to the fixer.
\subsubsection{Vulnerability Fixer (LLM) and Controlled Restart}
Attempts a constrained repair using all accumulated context (code, diffs, CWE, rationales, diagnostics). 
The repaired candidate re-enters validation at the \emph{Lightweight Static Analyzer} agent up to one time. 
If the intended vulnerability still does not materialize, the controller restarts from the initial \emph{Vulnerability Injector}, exploiting LLM stochasticity to explore a different edit trajectory. To ensure termination, we cap the total number of correction attempts (default: 10) across reinjection and refinement loops.

This hybrid validation strategy combines programmatic and rule-based tools, with LLM-based contextual reasoning, providing robust and interpretable guarantees over the correctness of injected vulnerabilities.

\section{Experimental Setup}
\label{sec:experimental_setup}

In this section, we detail the experimental design used to evaluate AVIATOR. We present our research questions, the datasets utilized for both injection validation and downstream tasks, and the implementation details of our framework. We aim to answer the following research questions:

\begin{itemize}
\item \textbf{RQ1:} How does AVIATOR perform compared to existing automated methods for creating realistic, large-scale vulnerability datasets?
\item \textbf{RQ2:} What is the contribution of fine-tuning the vulnerability injection agent to AVIATOR's overall performance?
\item \textbf{RQ3:} How do different compositions of agents (from monolithic to full agentic workflow) affect AVIATOR's vulnerability generation capabilities?
\item \textbf{RQ4:} How does AVIATOR's injection success rate varies across different weakness categories (CWEs)?
\item \textbf{RQ5:} Can AVIATOR-augmented datasets improve the performance of state-of-the-art deep learning-based vulnerability detection models compared to existing augmentation techniques?
\end{itemize}

\subsection{Datasets}
\label{subsec:datasets}
We evaluate AVIATOR specifically for C/C++ functions, given the inherent susceptibility of these languages to low-level vulnerabilities (e.g., memory errors), focusing on the function level in alignment with prior vulnerability detection and repair studies \cite{CppUnsafe_erlingsson2025}. 
For RQ1 to RQ4 we used three complementary popular open-source datasets used in previous studies \cite{sard100_hoole2016, formai_tihanyi2023, PrimeVul_ding2025}. For RQ5, we additionally used \cite{devign_zhou2019, bigvul_fan2020, PrimeVul_ding2025} following the methodology established in VulScribeR \cite{VulScribeR_2025}. Table~\ref{tab:datasets} summarizes the statistics of each dataset in our experiments.
\begin{table}[h]
    \centering
    \caption{Statistics of the Studied Datasets}
    \label{tab:datasets}
    \resizebox{\columnwidth}{!}{%
    \begin{tabular}{l|c|c|c|c|c|c}
        \toprule
        \textbf{Dataset} & \textbf{\makecell{Vulnerable\\Samples}} & \textbf{\makecell{Clean\\Samples}} & \textbf{Total} & \textbf{Ratio} & \textbf{CWE Num} & \textbf{Entropy} \\
        \midrule
        SARD100 \cite{sard100_hoole2016} & 102 & 102 & 204 & 1:1 & 23 & 3.98 \\
        \midrule
        FormAI \cite{formai_tihanyi2023} & 192,712 & 117,774 & 310,486 & 1:0.6 & 19 & 3.61 \\
        \midrule
        PrimeVul Train \cite{PrimeVul_ding2025} & 4,802 & 169,345 & 174,147 & 1:35.3 & 122 & 4.76 \\
        PrimeVul Validation & 593 & 23,347 & 23,940 & 1:39.4 & 73 & 5.05 \\
        PrimeVul Test & 549 & 24,232 & 24,781 & 1:44.1 & 71 & 5.09 \\
        \midrule
        Devign \cite{devign_zhou2019} & 10,768 & 12,024 & 22,792 & 1:1.1 & NA & 4.84 \\
        \midrule
        BigVul Train \cite{bigvul_fan2020} & 8,783 & 142,125 & 150,908 & 1:16.2 & 91 & 4.94 \\
        BigVul Validation & 1,038 & 17,826 & 18,864 & 1:17.2 & 84 & 4.95 \\
        BigVul Test & 1,079 & 17,785 & 18,864 & 1:16.5 & 85 & 4.72 \\
        \bottomrule
    \end{tabular}%
    }
\end{table}
\paragraph*{Entropy}
We report an embedding-based entropy score as a proxy for code diversity and richness, complementary to CWE count (which captures label-space coverage). Following~\cite{VulScribeR_2025}, we embed each function with CodeBERT, reduce embeddings to three dimensions using principal component analysis, discretize the space into a 10-bin histogram, and compute Shannon entropy over the resulting histogram. Higher entropy indicates a broader variety of code structures and idioms in the dataset, whereas lower entropy reflects more homogeneous or template-like code.

\begin{itemize}
\item \textit{SARD test suite 100} \cite{nist_sard_2018, sard100_hoole2016}: Contains 102 compact, synthetic test cases, including 34 vulnerable/benign function pairs that can be detected through formal verification tools such as ESBMC \cite{esbmc_gadelha2018}. Their compact and self-contained design, makes them ideal for rapid, fully automated validation and served as our primary development benchmark.
\item \textit{FormAI} \cite{formai_tihanyi2023}: 
Includes more realistic and structurally complex synthetic examples. Since it does not provide aligned benign/vulnerable pairs, we manually created benign versions for 38 randomly sampled vulnerable functions, without altering the program's intent and ensuring no vulnerabilities were detectable by formal verification tools on the resulting benign versions. ESBMC is again used for automated validation, offering a stronger challenge to our injection framework.
\item \textit{PrimeVul}~\cite{PrimeVul_ding2025}: A large dataset of more complex real-world code samples. Due to the lack of full context (e.g., missing class definitions, global variables) and larger code size, formal verification tools like ESBMC are not directly applicable. We use: (1) the paired training split with 3,789 corresponding pairs across 112 CWEs to fine-tune the Vulnerability Injector, (2) a subset of the full training set to train downstream DLVD, (3) the validation split of 480 corresponding pairs (63 CWEs) as the retrieval knowledge base in AVIATOR's RAG setup, (4) \al{100} randomly sampled pairs from the test split (\al{30} CWEs) for manual evaluation of injection success, and (5) the full test split (71 CWEs) to evaluate downstream DLVD performance.
\item \textit{Devign} \cite{devign_zhou2019}: Following previous studies~\cite{vgx_nong2024, vulgen_nong2023, VulScribeR_2025}, we use Devign as one of the two primary training datasets for the downstream tasks in RQ5. It offers a relatively balanced distribution between vulnerable and clean samples, making it suitable for training robust detection models without excessive class imbalance issues.
\item \textit{BigVul} \cite{bigvul_fan2020}: A large-scale real-world dataset. Consistent with the methodology in VulScribeR \cite{VulScribeR_2025}, we utilize the testing set of BigVul as an independent test set to evaluate the generalization capability of the vulnerability detection models trained on AVIATOR-augmented data (RQ5).
\end{itemize}

\subsection{Implementation Details}
\label{subsec:implementation}

\paragraph*{Agents}
AVIATOR leverages various technologies and tools. Within the validation workflow, we utilized the static analysis tool \textit{Cppcheck} version 2.17, due to its popularity and ease of integration.

We experimented with two base large language models for the AI agents in our workflow. Our default backbone is \textit{Qwen2.5-Coder-32B-Instruct}~\cite{qwen25coder_hui2024}, a specialized open-source coding LLM with 32 billion parameters. We additionally experimented with \textit{Llama-4-Maverick}~\cite{llama4_meta2025}, a general-purpose mixture-of-experts LLM with approximately 400 billion parameters (17 billion active parameters), to examine the impact of model specialization. \al{Unless otherwise stated, all experiments use the default coding backbone; Llama-4-Maverick is considered only in the model comparison of RQ3.}
\paragraph*{Retrieval-Augmented Generation}
We use \textit{gte-Qwen2-1.5B-Instruct} as the embedding model for vulnerable/benign code function pairs drawn from the validation split of PrimeVul (§\ref{subsec:datasets}) that constitutes the RAG knowledge base. At the time of our experiments, this model achieved state-of-the-art scores on the MTEB benchmark for embeddings up to 8K tokens~\cite{mteb_huggingface2023}. Since 99\% of functions in the PrimeVul training split fall below this length, no chunking was required. We set the retrieval parameter to $k=4$, following the recommended default.  

The \emph{Vulnerability Description} used by the \emph{Information Loader} agent (§\ref{sussubsec:information_loader}) covers 133 CWE categories and explains common programming mistakes that lead to each vulnerability type.
\paragraph*{Fine-Tuning Configurations}
We fine-tune the vulnerability injection agent using 3,789 aligned pairs across 112 CWEs from the PrimeVul training set.
For supervised fine-tuning (SFT), we trained for 5 epochs (loss convergence) with a learning rate of $2 \times 10^{-4}$, batch size of 4, and LoRA ranks varying between 16 and 256. 
For reinforcement learning with GRPO, we trained the model for 1 epoch (reward convergence), using a learning rate of $2 \times 10^{-4}$, batch size of 8, and generating 4 completions per policy iteration. The KL regularization coefficient $\beta$ (0.04) and clipping parameter $\epsilon$ (0.2) followed the recommended default values from GRPO implementation using Adam optimizer.
\paragraph*{Technology}
AVIATOR is implemented using Python as well as Hugging Face's Transformers and Parameter-Efficient Fine-Tuning libraries, and vLLM for inference. All experiments were run on NVIDIA A100 or H100 80GB GPUs.

\subsection{Empirical Study Design}

We conduct an empirical study to quantify the contribution of adding agentic components to the AVIATOR workflow. The goal is to assess both the necessity and the impact of the modular structure by systematically simplifying the framework and comparing outcomes. To this end, we construct and evaluate six progressively reduced variants of AVIATOR, obtained by incrementally removing agents from the full pipeline.

The complete AVIATOR pipeline presented in Figure~\ref{fig:aviator_architecture} showcases 13 distinct agents. These agents include tool-based nodes, LLM agents, and control logic designed to manage iterative processes.

The evaluated configurations are as follows:

\begin{itemize}
    \item \textbf{1-Agent (Monolithic LLM), $W_1$}: A single LLM receives the original function, vulnerability type, and retrieved examples, and is prompted to analyze the function, inject the vulnerability, and self-verify. This setting relies on common chain-of-thought prompting without agentic decomposition.

    \item \textbf{3-Agent, $W_3$}: Decomposes the task into three distinct agents: (i) loading the vulnerability description, (ii) analyzing the function for injection points, and (iii) performing the injection. No verification or refinement stages are included.

    \item \textbf{5-Agent (Injection-Only Module), $W_5$}: Implements the full Injection Module (Code Understanding, Information Loader, Vuln.-Specific Code Understanding, Vulnerability Injector, Polisher) without validation (Figure~\ref{fig:aviator_architecture}, top).

    \item \textbf{7-Agent, $W_7$}: Adds the difference-checking loop, consisting of the Diff Agent and Diff Analyzer, allowing the system to detect and iteratively refine the code if no substantive change was introduced.

    \item \textbf{9-Agent, $W_9$}: Adds the static analysis tool and corresponding Verifier Agent to further validate the generated vulnerability and refine injection quality based on external tool feedback.

    \item \textbf{11-Agent, $W_{11}$}: Incorporates the Diff Refiner agent after the difference is computed, allowing targeted edits based on diff feedback. This step improves alignment between injection goals and actual edits by making minimal context-aware adjustments.

    \item \textbf{13-Agent (Full AVIATOR), $W_{13}$}: Adds a terminal Vulnerability Fixer Agent. This final pass ensures that injected vulnerabilities are realistic, contextually valid, and non-trivially removable—closely mimicking the post-injection review performed by security experts.
\end{itemize}

All configurations share the same injection model, prompt formatting, and RAG setup. We evaluate these variants on the same benchmark split (see Section~\ref{subsec:datasets}) and report both formal and manual success metrics (see Section~\ref{subsec:eval_metrics}). This incremental comparison enables a fine-grained understanding of how individual agents affect the correctness, realism, and efficiency of the injection process.

\subsection{Evaluation and Metrics}
\label{subsec:eval_metrics}

\altwo{We use complementary metrics for the two evaluation objectives considered in this study. For RQ1-RQ4, we assess vulnerability injection success, as it directly indicates the expected label accuracy of datasets produced by our framework. For RQ5, we report precision, recall, and $F_1$ of the downstream detectors, following the VulScribeR protocol~\cite{VulScribeR_2025} and enabling consistent comparison across augmentation baselines. The additional detector-oriented metrics introduced in PrimeVul~\cite{PrimeVul_ding2025} are not consistently reported across these baselines and are therefore not used in this comparison.}

\paragraph*{Automated Evaluation}
For compatible test cases \cite{nist_sard_2018} and \cite{formai_tihanyi2023}, we automatically measured vulnerability injection success using the formal verification tool \textit{ESBMC} v7.5 (Efficient SMT-based Bounded Model Checker), recognized for its effectiveness in detecting vulnerabilities in bounded code contexts \cite{esbmc_gadelha2018, Sw_verification_competition_beyer2023}. It implements state-of-the-art incremental Bounded Model Checking and $k$-induction proof-rule algorithms. Namely, if the solver reaches termination within a bound less than or equal to $k$, we can definitively prove the absence of software errors. We set the bound to 100 and timeout to 5 minutes.

Due to the probabilistic nature of LLMs used in AVIATOR, identical inputs may yield different outputs on repeated trials. Therefore, we generated 5 outputs for SARD100 and FormAI test cases. To rigorously evaluate AVIATOR's performance, we employed two metrics:
\begin{itemize}
\item \textbf{Average Injection Success Rate (AISR)}: Measures the proportion of generated vulnerabilities correctly injected and verified, averaged across multiple runs. Denote $s_i$ as the success rate in run $i$ over $n$ runs:
{\small
\begin{equation}
\label{eq:asr}
AISR_n = \frac{1}{n} \sum_{i=1}^{n} s_i, \quad \sigma = \sqrt{\frac{1}{n} \sum_{i=1}^{n}(s_i - \text{AISR})^2},
\end{equation}
}
where $\sigma$ denotes the empirical standard deviation.

\item \textbf{Pass@k} \cite{passk_chen2021}: Estimates the probability that at least one of $k$ independently sampled outputs is correct. If $c$ is the number of correct outputs among $k$ samples:
\[
\text{Pass@}k =
\begin{cases}
1 - \dfrac{\binom{n-c}{k}}{\binom{n}{k}},
    & \text{if } c \leq n-k, \\[4pt]
1,  & \text{otherwise}.
\end{cases}
\]
This metric captures the practical success likelihood under multi-sample inference.
\end{itemize}

These metrics provide insights into AVIATOR's reliability and practical utility, reflecting its robustness and stability in automated vulnerability injection tasks.

\altwo{\paragraph*{Statistical Analysis}
A difference in injection success between two configurations may reflect run-to-run variation. We therefore test each comparison for statistical significance. All configurations inject the same code samples, so their measurements are paired by sample: for each sample, we computed the fraction of runs in which its injection succeeded, which gives one paired observation per sample ($N=38$ on FormAI and $N=34$ on SARD100). Most injections succeed, so these per-sample rates are strongly skewed toward $1$. We therefore used the two-sided \textit{paired Wilcoxon signed-rank test}, a non-parametric test that operates on the ranks of the paired differences and makes no normality assumption. We use the significance level of $\alpha=0.05$: a difference is statistically significant when its $p$-value falls below this threshold.}

\paragraph*{Manual Evaluation}
For PrimeVul, where formal verification tools such as ESBMC are not directly applicable due to the lack of full context (e.g., missing class definitions and global variables), as well as larger code samples, we manually validated a sample of \al{100} injected functions drawn uniformly at random, following prior work \cite{d2adatasetbuilt_zheng2021, DiverseVul_yizheng2023, vinj_nong2024, lessIsEnough_wen2023}. It is important to note that our total evaluation sample size is consistent with prior similar studies \cite{VulScribeR_2025, vinj_nong2024, d2adatasetbuilt_zheng2021}. The evaluation was performed by three authors experienced in C/C++, including two senior security researchers\al{, who independently analyzed each injected sample and assigned its final category by consensus}. Each injected sample was categorized as (1) weakness present, (2) no weakness, or (3) unknown, typically due to reliance on external functions or context beyond the injected function. \altwo{When the initial labels disagreed, the reviewers jointly reexamined the sample against the code and the target-CWE definition, with the two senior security researchers making the final decision; samples that remained ambiguous were labeled unknown.} Since we lack complete test cases for most programs, we assess the presence of weaknesses rather than confirmed vulnerabilities (which require demonstrable exploitability). The reported injection success rate reflects the percentage of samples found to contain a weakness, based only on those that could be conclusively assessed.

\subsection{Downstream Deep Learning-Based Vulnerability Detection (DLVD) Setup}
\label{subsec:downstream_setup}

To answer \textbf{RQ5}, we evaluate the practical utility of AVIATOR by using this workflow to augment training data for downstream vulnerability detection tasks. We adopt the experimental protocol established in recent state-of-the-art studies~\cite{VulScribeR_2025}, ensuring a fair comparison against existing augmentation techniques.

\subsubsection{Datasets and Augmentation Strategy}
We utilized two widely used benchmarks as seed datasets: \textit{Devign}~\cite{devign_zhou2019} and \textit{PrimeVul}~\cite{PrimeVul_ding2025}, described in Section~\ref{subsec:datasets}. For each dataset, we generated 5\,000 new vulnerable samples using AVIATOR and applied specific augmentation strategies to handle class balance, consistently with the methodology in \cite{VulScribeR_2025}:

\begin{itemize}
    \item \textbf{Devign Augmentation:} We added the 5\,000 AVIATOR-generated vulnerable samples to the training set. To prevent label shift and maintain the original class distribution, we simultaneously added a proportional number of benign samples. We sourced these benign samples from \cite{vgx_nong2024}, which have been previously deduplicated across datasets.
    
    \item \textbf{PrimeVul Augmentation:} Due to the extreme class imbalance in PrimeVul, we first down-sampled the benign samples in the training set to reduce the ratio from 1:35 to 1:15, making model training possible. After injecting 5\,000 AVIATOR-generated vulnerable instances, we applied Random Oversampling (ROS) exclusively to the original vulnerable samples to achieve a balanced 1:1 class ratio for training, to conduct a fair comparison between the methods \cite{VulScribeR_2025}.
\end{itemize}

For both datasets, we preserved the original distributions of the test sets to ensure that the evaluation assesses performance against real-world distributions.

\subsubsection{Models}
We evaluated performance across four distinct deep learning architectures, representing two major families of vulnerability detection models:

\begin{itemize}
    \item \textbf{Graph-based Models:} We employed \textit{Devign}~\cite{devign_zhou2019} and \textit{Reveal}~\cite{DLvulnDetectStudy_chakraborty2020}, which utilize Graph Neural Networks (GNNs) on Code Property Graphs (CPGs) to capture structural semantics.
    \item \textbf{Transformer-based Models:} We used \textit{LineVul}~\cite{linevul_fu2022}, a BERT-based model optimized for line-level detection, and \textit{CodeT5}~\cite{codet5_wang2021}, an encoder-decoder model pre-trained on code, used as a strong baseline in recent benchmarks~\cite{PrimeVul_ding2025}.
\end{itemize}

\textbf{Training Configuration:} To ensure a rigorous comparison, we adhered to the training settings reported in \cite{VulScribeR_2025}. For the graph-based models (Devign, Reveal), we trained each model 5 times with different random seeds and reported the results from the run achieving the highest $F_1$-score. For Transformer-based models (LineVul, CodeT5), we trained for 10 epochs and selected the checkpoint that achieved the highest $F_1$-score.

\subsubsection{Baselines}
We benchmarked AVIATOR against five distinct baselines\al{, including three vulnerability injection methods (VulGen, VGX, and VulScribeR)}:
\begin{enumerate}
    \item \textbf{No Augmentation (No Aug):} Training on the original dataset, to establish a reference.
    \item \textbf{Random Oversampling (ROS):} A standard class-balancing technique that duplicates vulnerable samples to mitigate imbalance, shown to be a competitive baseline in prior studies~\cite{ros_yang2023}.
    \item \textbf{VulGen}~\cite{vulgen_nong2023} and \textbf{VGX}~\cite{vgx_nong2024}: State-of-the-art injection techniques that rely on pattern mining and deep learning-based localization.
    \item \textbf{VulScribeR}~\cite{VulScribeR_2025}: A state-of-the-art generative augmentation method that uses LLMs and RAG to produce vulnerable samples, with results reported primarily on downstream DLVD performance.
\end{enumerate}

\section{Results and Analysis}
\label{sec:results}
In the following we investigate the answers to the research questions posed earlier. \al{Our evaluation assesses the quality of AVIATOR-generated datasets along two complementary axes: injection fidelity (RQ1--RQ4), which measures whether the generated samples are correctly labeled, and downstream detection utility (RQ5), which measures whether they improve real-world vulnerability detection. AVIATOR is optimized for neither metric directly; instead, both reflect the dataset quality its workflow is designed to deliver.}

\subsection{RQ1. Effectiveness of AVIATOR Compared to Existing Dataset Creation Techniques}

To evaluate AVIATOR's effectiveness, we assessed its vulnerability injection success using samples from three benchmarks: \textit{SARD100}, \textit{FormAI}, and \textit{PrimeVul}. Table~\ref{tab:rq1_results} shows the average success rates for the full AVIATOR workflow ($W_{13}$) with the SFT injection agent. 

AVIATOR achieved a 95\% success rate on SARD100 and 91\% on the more complex FormAI benchmark, indicating strong generalization beyond simple patterns. For PrimeVul, manual analysis was conducted on \al{100} injected samples due to the lack of external context needed for automated verification. Out of \al{97} analyzable cases, \al{90} contained a verifiable weakness within the function body (\al{93\%}). The remaining samples involved undefined external symbols, so the analysis focused on confirming the presence of weakness insertion rather than evaluating full vulnerability exploitability.

\begin{table}[h]
    \centering
    \caption{AVIATOR injection success across datasets: Average Injection Success Rate on 5 runs for SARD100 and FormAI, and on analyzable cases for PrimeVul due to limited context.}
    \label{tab:rq1_results}
    \resizebox{\linewidth}{!}{%
    \begin{tabular}{l|ccc}
        \hline
        \textbf{Dataset} & \textbf{Eval. Type} & \textbf{Metric} & \textbf{AVIATOR} \\
        \hline
        SARD100  & Automated & $AISR_5$ & 95\% \\
        FormAI   & Automated & $AISR_5$ & 91\% \\
        PrimeVul & Manual    & \makecell{\% Analyzable \\ Weaknesses} & \al{93\%} \\
    \end{tabular}}
\end{table}

Table~\ref{tab:benchmark} compares AVIATOR with other techniques, showing that its success rate across all three evaluation benchmarks (93\%) surpasses prior methods. \al{Following the cross-study comparison methodology of VulGen~\cite{vulgen_nong2023}, PrimeVul~\cite{PrimeVul_ding2025}, and VGX~\cite{vgx_nong2024}, we report each baseline's published figure (see the Source column). These figures are computed on data drawn from overlapping corpora that largely derive from the same CVE/NVD fixing-commit datasets~\cite{bigvul_fan2020,CrossVul_georgios2021,CVEfixes_guru2021}. A different controlled comparison on identical seed and test datasets is provided separately in RQ5 (Section~\ref{subsec:rq5_results}).} 

Earlier injection methods such as VulGen and VinJ that combine pattern mining with learned localization, report an injection success of 69\%, implying that a non-negligible fraction of generated samples do not instantiate the intended weakness. VGX follows a similar pattern-mining paradigm and substantially improves injection success (90\%), yet does so under tighter constraints, limited CWE coverage (23 types) and a primary focus on single-statement edits, which restrict the diversity of the vulnerabilities it can generate and, in turn, the downstream benefit of its data. In parallel, VulScribeR, an LLM- and RAG-based method whose strength is downstream augmentation, reports a lower estimated injection success (82\%). In contrast, AVIATOR's guided, multi-step injection workflow enables precise, context-aware edits that more closely resemble developer-introduced vulnerabilities, allowing it to remain effective even in partially specified or real-world code settings.

\begin{table}[h]
    \centering
    \caption{Label accuracy of major vulnerability datasets \al{and injection methods} satisfying the BCD properties (Section~\ref{subsec:properties_high_qual_vuln_dataset}). \al{Unmarked rows report audited label accuracy, whereas injection methods (*) report injection success, which by construction is the label accuracy of a dataset built by injection.} Results for AVIATOR are obtained via automated evaluation on SARD100 and FormAI, and manual validation on PrimeVul. The reported score for AVIATOR corresponds to the mean injection success rate across all three evaluation datasets.}
    \label{tab:benchmark}
    \begin{tabular}{l|cc}
        \hline
        \textbf{Model (*) / Dataset} & \textbf{Accuracy} & \textbf{Source} \\
        \hline
        Graph2Edit * \cite{Graph2Edit_yao2021} & 13\% & \cite{vulgen_nong2023} \\
        BigVul \cite{bigvul_fan2020} & 25\% & \cite{PrimeVul_ding2025} \\
        AVIATOR $W_1$ * & 31\% & FormAI validation \\
        CrossVul \cite{CrossVul_georgios2021} & 48\% & \cite{PrimeVul_ding2025} \\
        Getafix * \cite{Graph2Edit_yao2021} & 50\% & \cite{vulgen_nong2023} \\
        CVEFixes \cite{CVEfixes_guru2021} & 52\% & \cite{PrimeVul_ding2025} \\
        DiverseVul \cite{DiverseVul_yizheng2023} & 60\% & \cite{PrimeVul_ding2025} \\
        VulGen * \cite{vulgen_nong2023} & 69\% & \cite{vulgen_nong2023} \\
        VinJ * \cite{vinj_nong2024} & 69\% & \cite{vinj_nong2024} \\
        VulScribeR * \cite{VulScribeR_2025} & 82\% & \cite{VulScribeR_2025} \\
        VGX * \cite{vgx_nong2024} & 90\% & \cite{vgx_nong2024} \\
        AVIATOR $W_{13}$ \& SFT * & \textbf{93\%} & Cross-Dataset Eval. \\
    \end{tabular}
\end{table}

\begin{center}
\begin{tcolorbox}[myrqstyle]
\textbf{Answer to RQ1:} AVIATOR achieves a high injection success across benchmarks: 95\% on \textsc{Sard100}, 91\% on \textsc{FormAI}, and \al{93\%} verifiable weakness insertion on \textsc{PrimeVul}. These results highlight AVIATOR's strong generalization, even in partially specified or real-world code settings. We attribute this performance to AVIATOR's novel use of an agentic workflow that incorporates multi-step guidance and task decomposition, which enables the injection agent to iteratively refine and validate its own output, likely contributing to its effectiveness across diverse benchmarks.
\end{tcolorbox}
\end{center}

\subsection{RQ2. Impact of Fine-Tuning the Injection Agent on AVIATOR's Performance}

To assess the effect of model-level adaptation, we evaluate AVIATOR's injection performance before and after fine-tuning the injection agent. We experiment with two fine-tuning regimes: supervised fine-tuning (SFT) and reinforcement learning via Group Relative Policy Optimization (GRPO), using the same default backbone and the full agentic pipeline ($W_{13}$).

Crucially, all fine-tuning is performed exclusively on PrimeVul, a dataset that shares no overlap with either SARD100 or FormAI. This ensures the evaluation measures true generalization, not memorization.

\begin{table}[h]
  \centering
  \caption{AVIATOR $W_{13}$ injection success rate ($AISR_5$) across five vulnerability injection runs with and without fine-tuning, on FormAI and SARD100.}
  \label{tab:rq2_results}
  \begin{tabular}{l|cc|cc}
      \hline
      \textbf{Dataset} & \textbf{Model Variant} & \textbf{$AISR_5$} & \textbf{Std} \\
      \hline
      \multirow{3}{*}{FormAI} 
      & Base model & 85\% & 4 \\
      & + GRPO & 84\% & 1 \\
      & + SFT & \textbf{91\%} & 1 \\
      \hline
      \multirow{3}{*}{SARD100} 
      & Base model & 94\% & 3 \\
      & + GRPO & 94\% & 0 \\
      & + SFT & \textbf{95\%} & 2 \\
      \hline
  \end{tabular}
\end{table}

\begin{figure}[h]
  \begin{tabular}{cc}
       \includegraphics[width=0.5\linewidth]{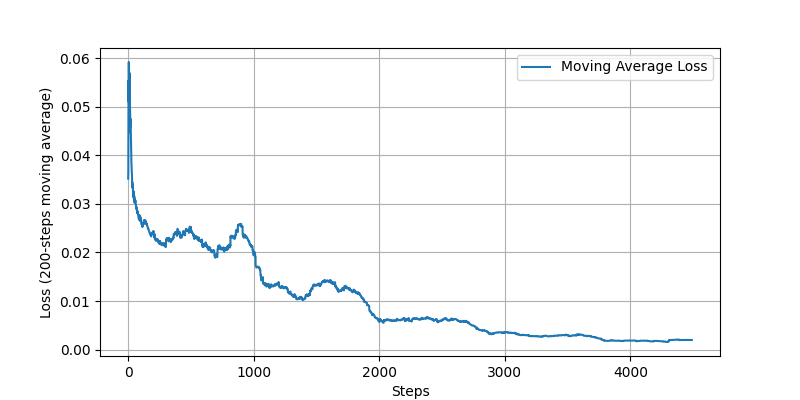}
&
       \includegraphics[width=0.5\linewidth]{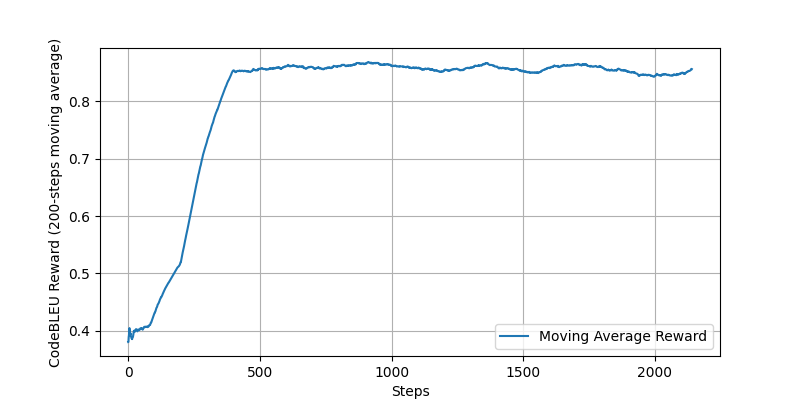}
   \end{tabular}
   \caption{Training dynamics of the vulnerability injection agent. Left: SFT training loss over 5 epochs, with lr=$2 \times 10^{-4}$, batch size = 4, LoRA $r=32$, and $\alpha=32$. Right: GRPO CodeBLEU reward over 2 iterations, with lr=$2 \times 10^{-4}$, batch size = 8, 4 generated samples per policy iteration, LoRA $r=32$, and $\alpha=32$.}
   \label{fig:training_metrics}
\end{figure}

\paragraph*{SFT is both more effective and more efficient than GRPO}
Table~\ref{tab:rq2_results} reports the average injection success rate ($AISR_5$) and its standard deviation across 5 vulnerability injection runs on the same data. On the more complex FormAI dataset, SFT yields significant gains over the base model (91\% vs. 85\%\al{; $p=0.049$}) \al{and over GRPO ($p=0.041$, two-sided paired Wilcoxon signed-rank test over per-sample success rates; Table~\ref{tab:rq2_wilcoxon})}\altwo{, both below the $\alpha=0.05$ threshold}, while also reducing variance. \althree{These near-threshold $p$-values are further corroborated by converging evidence: both comparisons agree in direction, variance is reduced, and Pass@$k$ improves consistently (Table~\ref{tab:combined_passatk}).} In contrast, GRPO not only fails to outperform the base model on FormAI (dropping to 84\%\altwo{; not significant, $p=0.69$}), but also provides no gain on SARD100. 
\altwo{The early saturation of the GRPO reward (Figure~\ref{fig:training_metrics}, right) suggests that further CodeBLEU optimization provides limited additional learning signal, which we analyze further below.} Overall, these results indicate that SFT provides a more stable and effective optimization strategy for vulnerability injection in AVIATOR.
On the simpler SARD100 benchmark, the gains from SFT are marginal (95\% vs. 94\%\al{, not significant}\altwo{; all three comparisons on this benchmark yield $p \geq 0.46$, well above $\alpha=0.05$}), indicating that AVIATOR's base configuration already performs near-saturation.

\begin{table}[h]
\centering
\caption{Paired Wilcoxon signed-rank test on per-sample success rates across 5 injection runs. Significant results ($p<0.05$) are shown in bold.}
\label{tab:rq2_wilcoxon}
\begin{tabular}{l|l|c}
    \hline
    \textbf{Dataset} & \textbf{Comparison} & \textbf{$p$-value} \\
    \hline
    \multirow{3}{*}{FormAI ($N=38$)}
    & SFT vs.\ no FT  & \textbf{0.049} \\
    & SFT vs.\ GRPO   & \textbf{0.041} \\
    & GRPO vs.\ no FT & 0.69 \\
    \hline
    \multirow{3}{*}{SARD100 ($N=34$)}
    & SFT vs.\ no FT  & 0.63 \\
    & SFT vs.\ GRPO   & 1.00 \\
    & GRPO vs.\ no FT & 0.46 \\
    \hline
\end{tabular}
\end{table}

\paragraph*{Pass@k Improvements with SFT}
We also reported Pass@k performance to assess the likelihood of producing a successful injection within $k$ generation attempts. Table~\ref{tab:combined_passatk} shows that SFT improves results consistently across all $k$ on both benchmarks. On FormAI, Pass@1 improves by over 5 percentage points (from 84.3\% to 89.9\%), and SARD100 reaches near-perfect performance even for small $k$. These gains further validate the robustness added by SFT.

\begin{table*}[h]
\centering
\caption{AVIATOR $W_{13}$ pass@k for 10 vulnerability injection generations with and without fine-tuning, on FormAI and SARD100.}
\label{tab:combined_passatk}
\resizebox{\linewidth}{!}{%
\begin{tabular}{c|c|c|c|c|c|c|c|c|c|c|c}
\hline
\textbf{Dataset} & \textbf{Model} & @1 & @2 & @3 & @4 & @5 & @6 & @7 & @8 & @9 & @10 \\
\hline
\multirow{2}{*}{FormAI} 
& $W_{13}$ & 84.32\% & 91.05\% & 93.87\% & 95.24\% & 96.03\% & 96.56\% & 96.94\% & 97.18\% & 97.30\% & 97.30\% \\
& $W_{13}$ + SFT & \textbf{89.91\%} & \textbf{93.04\%} & \textbf{94.81\%} & \textbf{95.62\%} & \textbf{96.15\%} & \textbf{96.55\%} & \textbf{96.87\%} & \textbf{97.14\%} & \textbf{97.38\%} & \textbf{97.58\%} \\
\hline
\multirow{2}{*}{SARD100} 
& $W_{13}$ & 93.58\% & 95.95\% & 96.66\% & 96.91\% & 97.01\% & 97.05\% & 97.06\% & 97.06\% & 97.06\% & 97.06\% \\
& $W_{13}$ + SFT & \textbf{95.15\%} & \textbf{98.65\%} & \textbf{99.51\%} & \textbf{99.79\%} & \textbf{99.93\%} & \textbf{99.99\%} & \textbf{100.00\%} & \textbf{100.00\%} & \textbf{100.00\%} & \textbf{100.00\%} \\
\hline
\end{tabular}}
\end{table*}

It is important to note that, even without SFT, AVIATOR achieves strong baseline performance. This indicates that the agentic workflow, through decomposition, correction loops, and validation, is sufficient to generate realistic and diverse vulnerabilities. Fine-tuning provides further gains but is not required for the framework to operate effectively. While baseline performance depends on how well the underlying LLM has been trained on the target language, this property remains especially valuable when extending to programming languages or vulnerability classes where aligned training data is scarce, since AVIATOR can still produce rich vulnerability patterns.

\paragraph*{Learning dynamics}
Figure~\ref{fig:training_metrics}, left, shows the SFT training loss, which demonstrates stable convergence within five epochs. In contrast, GRPO (Figure~\ref{fig:training_metrics}, right) achieves rapid early gains in CodeBLEU reward before plateauing. \altwo{One plausible explanation is that vulnerability injection typically modifies only a few security-critical statements while preserving most of the original function. Once candidates reach high structural similarity, CodeBLEU therefore provides limited discrimination between outputs that may still differ in whether the intended weakness is successfully introduced. A reward that directly captures successful weakness introduction could provide a stronger signal, but no automated oracle covers all supported CWE types, and running one at every rollout would require prohibitive compute.} GRPO also incurs substantially higher computational cost, requiring 10 days of training compared to only 10 hours for SFT.

\begin{center}
\begin{tcolorbox}[myrqstyle]
\textbf{Answer to RQ2.} Fine-tuning the vulnerability injection agent using SFT leads to consistent and significant performance gains for the whole AVIATOR workflow, particularly on complex datasets like FormAI. GRPO fails to match this improvement despite being substantially more resource intensive. These results highlight the efficiency and generalization strength of SFT, even when trained on an entirely separate dataset. 
Additionally, the results show strong baseline performance without fine-tuning, indicating that the agentic workflow alone can inject vulnerabilities. This is essential for transfer to new programming languages or CWE categories where aligned data is unavailable.
\end{tcolorbox}
\end{center}

\subsection{RQ3. Contribution of Agentic Components and Model Variants to Injection Performance}

\begin{figure*}[t!]
   \centering
\begin{tabular}{cc}  
\includegraphics[width=0.4\textwidth]{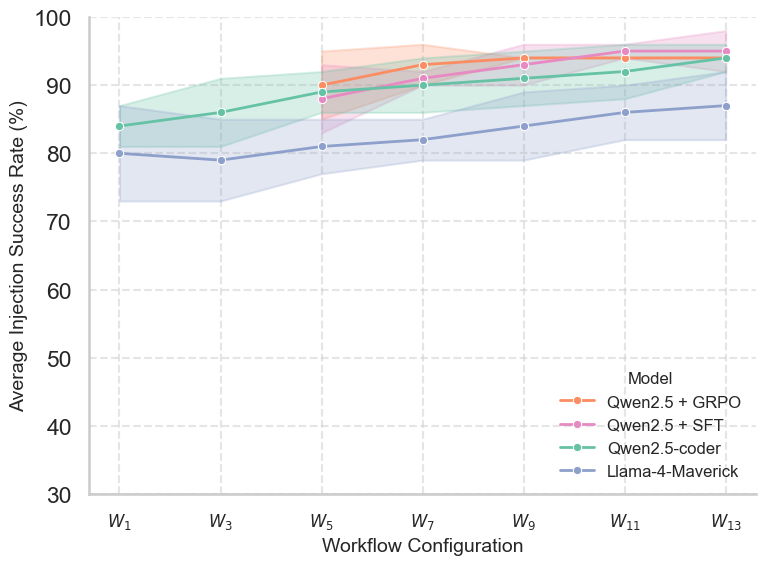}
&   
\includegraphics[width=0.4\textwidth]{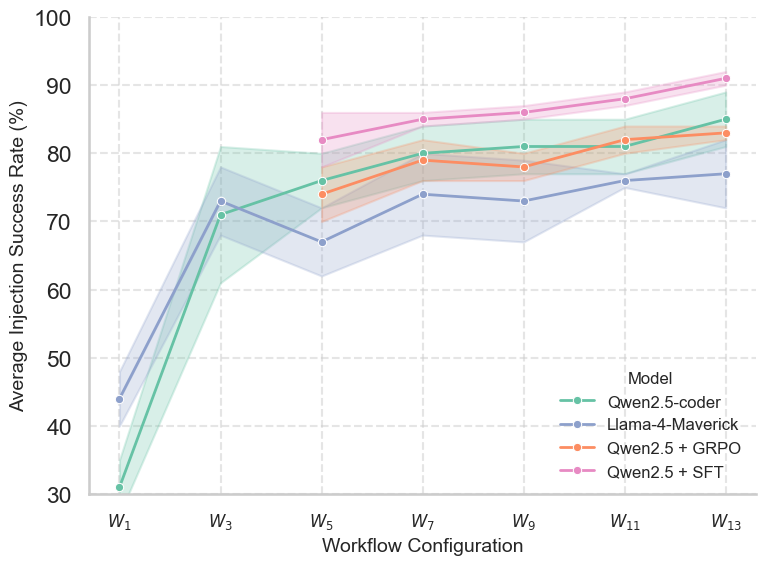}\\
  (a) SARD 100  & (b) FormAI
\end{tabular}
  \caption{AVIATOR average injection success rate on 5 attempts. Empirical study for LLM agents using Llama-4-Maverick, Qwen2.5-Coder with and without the vulnerability injection agent fine-tuned using SFT and GRPO.}
  \label{fig:sard100_ablation}
\end{figure*}

To evaluate the contribution of AVIATOR's agentic workflow design, we conducted an empirical study on seven progressively extended configurations ($W_1$–$W_{13}$), incrementally adding specialized agents for analysis, validation, and refinement. We additionally compare four model variants: (i) the base coding model, (ii) its SFT-adapted variant, (iii) its GRPO-adapted variant, and (iv) the general-purpose Llama-4-Maverick. Figure~\ref{fig:sard100_ablation} shows the average injection success rate across 5 trials ($AISR_5$) on both SARD100 and FormAI.

\paragraph*{Performance Improves with Workflow Depth}
Across all models, performance increases consistently as more specialized agents are added. For example, with the base model, success rises from 31\% at $W_1$ to 85\% at $W_{13}$ on FormAI, and from 84\% to 94\% on SARD100. The benefits of modular reasoning are particularly pronounced on complex tasks (FormAI), confirming that decomposing the task into subtasks handled by agents enhances robustness and clarity.

\paragraph*{Impact of Specific Agentic Additions}
Moving from $W_1 \rightarrow W_3$ introduces basic task decomposition, yielding large but unstable gains on FormAI (+40\%) due to better prompt structuring.
At $W_5$, AVIATOR includes the complete injection pipeline, stabilizing outputs and significantly improving accuracy (+5-10\%). Furthermore, adding feedback-based validation via diff checking ($W_7$) and static analysis with \texttt{Cppcheck} ($W_9$) further boosts performance, particularly for models with more specialized abilities like the fine-tuned version. Final refiners ($W_{11}$, $W_{13}$) provide consistent final gains across models, enabling subtle bug integration and polishing.

\paragraph*{SFT and GRPO Validated Across Configurations}
To ensure consistent comparisons, we fine-tuned the vulnerability injection agent starting from $W_5$, where the agent's input structure and prompt format remain unchanged across workflows. Configurations below $W_5$ ($W_1$, $W_3$) are not compatible with our fine-tuned models, as they would require new training due to different agent prompts and inputs. Consistent with results in RQ2, SFT delivers the highest gains across all configurations, while GRPO fails to improve over the base model despite significantly longer training time.

\paragraph*{General-Purpose vs. Specialized Models} Llama-4-Maverick, despite its substantially larger parameter count (400B), underperforms the 32B specialized coding backbone and its adapted variants. For example, on FormAI at $W_{13}$, it reaches 77\%, compared with 85\% for the base coding model and 91\% after SFT. This underlines the advantage of coding-specialized models and domain-specific fine-tuning for structured tasks like vulnerability injection.

\altwo{Appendix~\ref{appendix:model_sizes} extends this comparison to smaller models of the Qwen2.5-Coder family, from 1.5B to 14B parameters, and confirms that the gains from the agentic decomposition hold at every model size.}

\begin{center}
\begin{tcolorbox}[myrqstyle]
\textbf{Answer to RQ3.} Injection performance improves with workflow refinement: from 31\% at $W_1$ to 91\% at $W_{13}$ on FormAI. Fine-tuning the injection agent using SFT amplifies these gains even though most of the base performance comes from a well-crafted workflow. GRPO fails to outperform the base model. Domain-specialized coding models clearly outperform the larger general-purpose model considered in our experiments, demonstrating the complementary value of model specialization and expert-based agentic decomposition.
\end{tcolorbox}
\end{center}


\subsection{RQ4. Performance Variability Across Vulnerability Categories}
\label{subsec:rq4_variability}

\begin{figure*}[ht]
    \centering
    \includegraphics[width=0.9\linewidth]{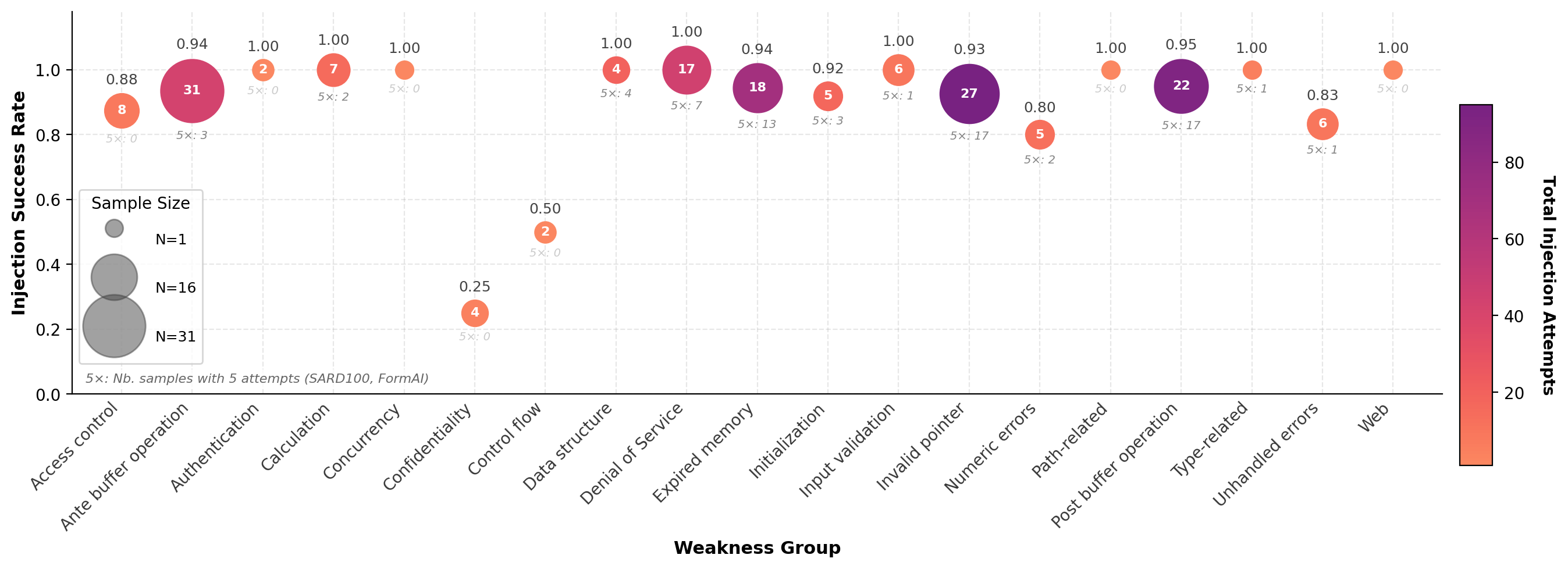}
    \caption{Injection success rate by \altwo{weakness group ~\cite{sate5_delaitre2018}} within the evaluation datasets. Bubble size, and the value annotated inside it, indicate the number of unique code samples in the group. The annotation \textit{5×} denotes samples from SARD100 and FormAI evaluated across 5 repeated runs; their reported success rate is the average over the 5 runs. The color gradient reflects the total volume of injection attempts, accounting for all repetitions.}
    \label{fig:group_success_distribution}
\end{figure*}

To further characterize the capabilities of AVIATOR, we analyze injection success rates across the CWE categories present in our evaluation benchmarks (SARD100, FormAI, and the PrimeVul subset). \altwo{Individual CWEs are each represented by only a few samples, so we report success by weakness group, using the grouping of the SATE V report~\cite{sate5_delaitre2018}; the per-CWE breakdown is given in Appendix~\ref{appendix:cwe_success}.} As shown in Figure~\ref{fig:group_success_distribution}, this distribution reflects only the categories present in the datasets; the current AVIATOR instantiation supports 133 CWE types in total as tested in \ref{subsec:rq5_results}.

Across the evaluated benchmarks, AVIATOR achieves a high overall injection success rate of 93\%. In particular, the framework performs strongly on vulnerability types listed in the MITRE Top 25 Most Dangerous Software Weaknesses. This suggests that AVIATOR is well-aligned with real-world security priorities and effectively injects vulnerabilities that matter most in practice.

At a finer granularity, AVIATOR performs well on both simple and complex vulnerability injections. It reliably injects vulnerabilities realizable through localized edits (e.g., missing checks or unsafe memory accesses), while also succeeding on categories that require non-trivial reasoning over program state and execution order. For instance, AVIATOR achieves high success rates on \textbf{CWE-416} (Use After Free, \altwo{$N=12$}), \textbf{CWE-476} (NULL Pointer Dereference, \altwo{$N=27$}), and \textbf{CWE-120} (Buffer Copy without Checking Size of Input, $N=12$), which require consistent reasoning about data flow, pointer validity, and memory access constraints. Beyond memory safety, AVIATOR also performs reliably on arithmetic and control-dependent vulnerabilities such as \textbf{CWE-369} (Divide by Zero, $N=5$), demonstrating versatility across vulnerability classes.

However, Figure~\ref{fig:group_success_distribution} also reveals performance variability primarily on vulnerability categories that either lack precise injection semantics or require constraint satisfaction beyond the function scope. \altwo{The Confidentiality group achieves a 25\% success rate and gathers CWEs that, although still widely used in vulnerability records, MITRE recommends deferring to the more specific weakness that causes the disclosure~\cite{cwe_mitre2026}, making it difficult to formulate a concrete injection objective. The Control flow group, at 50\%, covers only two samples, and its failing case requires the affected program point to be reachable by an attacker, a condition that depends on the callers of the function and on external inputs. These cases reflect inherent limitations of function-scoped injections. All groups with at least five samples reach a success rate of 0.80 or above. Re-injecting the four samples of the Confidentiality group with a specific weakness in place of the high-level CWE, raises injection success from 25\% to 75\% (Appendix~\ref{appendix:deferral})}.

\begin{center}
\begin{tcolorbox}[myrqstyle]
\textbf{Answer to RQ4:} AVIATOR achieves a high overall injection success rate of 93\% and performs strongly on practically relevant vulnerabilities, including multiple CWEs from the MITRE Top 25. The framework successfully injects non-trivial vulnerabilities that require reasoning about data flow, control flow, and object lifetimes (e.g., CWE-416, CWE-476), while lower success rates remain confined to ambiguous weakness categories, or to those requiring context beyond the function.
\end{tcolorbox}
\end{center}

\subsection{Compute and Cost Estimates}
On 8 NVIDIA A100 GPUs, running 128 jobs in parallel, we generated 5\,000 vulnerable samples in 4h37m. This averages to 3s per injection with concurrency. At Maryland's commercial electricity rate (14.82¢/kWh), this corresponds to a cost of about \textbf{\$0.44 per 1\,000 injections} (GPU power only). These numbers show that AVIATOR can effectively scale dataset generation at very low cost.
\altwo{\paragraph*{Low-resource deployment}
Serving the 32B model at 4 bits requires 25~GB and at 2 bits 16~GB, the latter running on a single 24~GB consumer GPU, with minimal loss of injection success and still above any smaller model of the same family that fits the same memory (Appendix~\ref{appendix:model_sizes}). Our run corresponds to about 7.4 GPU-hours per 1\,000 injections, so fewer GPUs trade throughput for running time, and a hosted endpoint removes the local GPU requirement altogether.}

\subsection{RQ5. Impact on Downstream Vulnerability Detection}
\label{subsec:rq5_results}

To assess AVIATOR's practical utility, we evaluate how augmenting training data with AVIATOR-generated vulnerable samples impacts downstream deep learning-based vulnerability detection (DLVD) performance. Table~\ref{tab:downstream_results} reports the performance metrics for four distinct deep learning architectures (Devign, Reveal, LineVul, CodeT5) trained on the \textit{Devign} and \textit{PrimeVul} datasets augmented with generated samples using different strategies.

\begin{table*}[ht]
    \centering
    \caption{Downstream vulnerability detection performance (Precision, Recall, $F_1$). Each model is trained on the seed dataset augmented with 5k injected vulnerable samples using each strategy. Best $F_1$ scores per model and setting are highlighted in \textbf{bold}. Results for non-AVIATOR strategies are taken from the VulScribeR paper \cite{VulScribeR_2025}; we report new results for AVIATOR and CodeT5.}
    \label{tab:downstream_results}
    \resizebox{\textwidth}{!}{%
    \begin{tabular}{l|ccc|ccc|ccc|ccc}
        \toprule
        \multirow{2}{*}{\textbf{Strategy}} & \multicolumn{3}{c|}{\textbf{Devign Model}} & \multicolumn{3}{c|}{\textbf{Reveal Model}} & \multicolumn{3}{c|}{\textbf{LineVul Model}} & \multicolumn{3}{c}{\textbf{CodeT5 Model}} \\
        \cmidrule{2-13}
         & P & R & $F_1$ & P & R & $F_1$ & P & R & $F_1$ & P & R & $F_1$ \\
        \midrule
        \multicolumn{13}{c}{\textit{Seed Dataset: Devign};  \textit{Test Dataset: BigVul Test}} \\
        \midrule
        No Aug
        & \heatcell{6.27}{6.06}{9.59} & \heatcell{43.24}{24.80}{62.32} & \heatcell{10.95}{10.95}{13.83}
        & \heatcell{7.28}{6.36}{11.64} & \heatcell{77.27}{33.39}{77.27} & \heatcell{13.31}{11.66}{17.82}
        & \heatcell{7.90}{5.35}{11.43} & \heatcell{29.19}{13.16}{98.82} & \heatcell{12.43}{10.15}{16.38}
        & \heatcell{6.08}{6.08}{8.08} & \heatcell{65.71}{31.05}{80.17} & \heatcell{11.13}{11.13}{14.39} \\
        ROS
        & \heatcell{7.54}{6.06}{9.59} & \heatcell{25.60}{24.80}{62.32} & \heatcell{11.65}{10.95}{13.83}
        & \heatcell{7.96}{6.36}{11.64} & \heatcell{33.39}{33.39}{77.27} & \heatcell{12.86}{11.66}{17.82}
        & \heatcell{10.91}{5.35}{11.43} & \heatcell{13.16}{13.16}{98.82} & \heatcell{11.93}{10.15}{16.38}
        & \heatcell{7.52}{6.08}{8.08} & \heatcell{51.81}{31.05}{80.17} & \heatcell{13.13}{11.13}{14.39} \\
        VulGen
        & \heatcell{6.49}{6.06}{9.59} & \heatcell{62.32}{24.80}{62.32} & \heatcell{11.75}{10.95}{13.83}
        & \heatcell{6.36}{6.36}{11.64} & \heatcell{70.59}{33.39}{77.27} & \heatcell{11.66}{11.66}{17.82}
        & \heatcell{8.51}{5.35}{11.43} & \heatcell{22.15}{13.16}{98.82} & \heatcell{12.29}{10.15}{16.38}
        & \heatcell{7.34}{6.08}{8.08} & \heatcell{31.05}{31.05}{80.17} & \heatcell{11.87}{11.13}{14.39} \\
        VGX
        & \heatcell{6.06}{6.06}{9.59} & \heatcell{59.30}{24.80}{62.32} & \heatcell{10.99}{10.95}{13.83}
        & \heatcell{6.89}{6.36}{11.64} & \heatcell{73.29}{33.39}{77.27} & \heatcell{12.60}{11.66}{17.82}
        & \heatcell{5.35}{5.35}{11.43} & \heatcell{98.82}{13.16}{98.82} & \heatcell{10.15}{10.15}{16.38}
        & \heatcell{7.20}{6.08}{8.08} & \heatcell{80.17}{31.05}{80.17} & \heatcell{13.21}{11.13}{14.39} \\
        VulScribeR
        & \heatcell{8.79}{6.06}{9.59} & \heatcell{29.41}{24.80}{62.32} & \heatcell{13.53}{10.95}{13.83}
        & \heatcell{11.64}{6.36}{11.64} & \heatcell{38.00}{33.39}{77.27} & \heatcellbf{17.82}{11.66}{17.82}
        & \heatcell{11.43}{5.35}{11.43} & \heatcell{18.81}{13.16}{98.82} & \heatcell{14.22}{10.15}{16.38}
        & \heatcell{7.09}{6.08}{8.08} & \heatcell{76.09}{31.05}{80.17} & \heatcell{12.97}{11.13}{14.39} \\
        AVIATOR
        & \heatcell{9.59}{6.06}{9.59} & \heatcell{24.80}{24.80}{62.32} & \heatcellbf{13.83}{10.95}{13.83}
        & \heatcell{9.19}{6.36}{11.64} & \heatcell{56.94}{33.39}{77.27} & \heatcell{15.82}{11.66}{17.82}
        & \heatcell{11.27}{5.35}{11.43} & \heatcell{29.99}{13.16}{98.82} & \heatcellbf{16.38}{10.15}{16.38}
        & \heatcell{8.08}{6.08}{8.08} & \heatcell{65.89}{31.05}{80.17} & \heatcellbf{14.39}{11.13}{14.39} \\
        \midrule
        \multicolumn{13}{c}{\textit{Seed Dataset: PrimeVul Train};  \textit{Test Dataset: PrimeVul Test}} \\
        \midrule
        No Aug
        & \heatcell{25.93}{4.94}{25.93} & \heatcell{5.12}{5.12}{59.76} & \heatcell{8.55}{8.55}{13.45}
        & \heatcell{7.35}{5.29}{8.65} & \heatcell{44.63}{36.32}{59.27} & \heatcell{12.62}{9.72}{13.97}
        & \heatcell{18.31}{17.29}{19.77} & \heatcell{31.15}{27.69}{31.88} & \heatcell{23.06}{22.42}{23.18}
        & \heatcell{19.34}{16.61}{20.26} & \heatcell{25.68}{24.04}{32.79} & \heatcell{22.07}{21.73}{22.44} \\
        ROS
        & \heatcell{7.45}{4.94}{25.93} & \heatcell{52.44}{5.12}{59.76} & \heatcell{13.04}{8.55}{13.45}
        & \heatcell{7.58}{5.29}{8.65} & \heatcell{43.66}{36.32}{59.27} & \heatcell{13.07}{9.72}{13.97}
        & \heatcell{18.48}{17.29}{19.77} & \heatcell{30.05}{27.69}{31.88} & \heatcell{22.89}{22.42}{23.18}
        & \heatcell{18.03}{16.61}{20.26} & \heatcell{28.42}{24.04}{32.79} & \heatcell{22.07}{21.73}{22.44} \\
        VulGen
        & \heatcell{4.94}{4.94}{25.93} & \heatcell{59.76}{5.12}{59.76} & \heatcell{9.11}{8.55}{13.45}
        & \heatcell{5.29}{5.29}{8.65} & \heatcell{59.27}{36.32}{59.27} & \heatcell{9.72}{9.72}{13.97}
        & \heatcell{19.39}{17.29}{19.77} & \heatcell{28.23}{27.69}{31.88} & \heatcell{22.99}{22.42}{23.18}
        & \heatcell{19.82}{16.61}{20.26} & \heatcell{24.04}{24.04}{32.79} & \heatcell{21.73}{21.73}{22.44} \\
        VGX
        & \heatcell{5.61}{4.94}{25.93} & \heatcell{53.66}{5.12}{59.76} & \heatcell{10.15}{8.55}{13.45}
        & \heatcell{5.44}{5.29}{8.65} & \heatcell{50.73}{36.32}{59.27} & \heatcell{9.82}{9.72}{13.97}
        & \heatcell{19.77}{17.29}{19.77} & \heatcell{27.69}{27.69}{31.88} & \heatcell{23.06}{22.42}{23.18}
        & \heatcell{16.61}{16.61}{20.26} & \heatcell{32.79}{24.04}{32.79} & \heatcell{22.05}{21.73}{22.44} \\
        VulScribeR
        & \heatcell{7.57}{4.94}{25.93} & \heatcell{55.37}{5.12}{59.76} & \heatcell{13.31}{8.55}{13.45}
        & \heatcell{7.81}{5.29}{8.65} & \heatcell{49.76}{36.32}{59.27} & \heatcell{13.51}{9.72}{13.97}
        & \heatcell{17.29}{17.29}{19.77} & \heatcell{31.88}{27.69}{31.88} & \heatcell{22.42}{22.42}{23.18}
        & \heatcell{20.26}{16.61}{20.26} & \heatcell{25.14}{24.04}{32.79} & \heatcellbf{22.44}{21.73}{22.44} \\
        AVIATOR
        & \heatcell{7.73}{4.94}{25.93} & \heatcell{51.82}{5.12}{59.76} & \heatcellbf{13.45}{8.55}{13.45}
        & \heatcell{8.65}{5.29}{8.65} & \heatcell{36.32}{36.32}{59.27} & \heatcellbf{13.97}{9.72}{13.97}
        & \heatcell{19.74}{17.29}{19.77} & \heatcell{28.05}{27.69}{31.88} & \heatcellbf{23.18}{22.42}{23.18}
        & \heatcell{16.96}{16.61}{20.26} & \heatcell{31.33}{24.04}{32.79} & \heatcell{22.01}{21.73}{22.44} \\
        \bottomrule
    \end{tabular}}
\end{table*}

\begin{table}[h] 
    \centering 
    \caption{Overall Average of Per-Configuration Relative Improvement of AVIATOR Compared to Other Strategies.} 
    \label{tab:avg_improvement} 
    \begin{tabular}{l|c|c|c} 
        \toprule 
        \textbf{Compared Strategy} & \textbf{Precision} & \textbf{Recall} & \textbf{$F_1$-Score} \\
        \midrule 
        No Aug & +12.2\% & +104.9\% & \textbf{+21.8\%} \\ 
        ROS & +9.0\% & +26.0\% & \textbf{+12.5\%} \\ 
        VulGen & +30.3\% & +5.7\% & \textbf{+25.2\%} \\ 
        VGX & +39.2\% & -25.4\% & \textbf{+24.6\%} \\ 
        VulScribeR & +1.4\% & +7.4\% & \textbf{+2.9\%} \\ 
        \bottomrule 
    \end{tabular} 
\end{table}

\paragraph*{Consistent downstream gains and comparison to augmentation baselines}
Table~\ref{tab:downstream_results} shows that AVIATOR consistently improves downstream DLVD performance across detector families (GNN-based Devign/Reveal and Transformer-based LineVul/CodeT5) and across evaluation settings (cross-dataset Devign$\rightarrow$BigVul and in-domain PrimeVul). Across the eight model-dataset configurations, AVIATOR achieves the best $F_1$ score in \textbf{6/8} cases. As further summarized in Table~\ref{tab:avg_improvement}, AVIATOR also delivers the \textbf{largest average $F_1$ improvement} among all augmentation strategies. This consistent performance across diverse architectures and datasets indicates that the data generated by AVIATOR exhibits strong generalizability and semantic richness, rather than being tailored to a specific detector family.

AVIATOR leads to an average $F_1$ score improvement of \textbf{+21.8\%} relative to the original dataset (\textbf{No Aug}), confirming that the injected samples provide substantial additional training signal beyond the original data. Compared to the Random Oversampling baseline (\textbf{ROS}), AVIATOR achieves a \textbf{+12.5\%} average $F_1$ score improvement. Notably, ROS itself outperforms VulGen and VGX by 9.0\% and 8.5\% in average $F_1$, respectively, highlighting the strength of AVIATOR beyond simple class-balancing effects.

When compared to pattern-based injection methods, AVIATOR improves average $F_1$ score by \textbf{+25.2\%} over \textbf{VulGen} and by \textbf{+24.6\%} over \textbf{VGX}. Among all baselines, VGX is the only method that achieves a higher average recall than AVIATOR; however, this comes at the cost of substantially lower precision. This behavior aligns with VGX's reliance on single-statement injection templates and its limited coverage of vulnerability classes (23 CWEs). Such augmentation strategies can make detectors overly sensitive, yielding high recall but low precision (many false positives) which ultimately limits their practical utility for large-scale data augmentation \cite{VulScribeR_2025}. In contrast, AVIATOR improves $F_1$ through a more balanced precision-recall profile, indicating that it provides a cleaner and less artifact-driven training signal.

Finally, AVIATOR outperforms \textbf{VulScribeR} by \textbf{+2.9\%} in average $F_1$ \althree{, achieving the best $F_1$ score in 6 of the 8 model--dataset configurations}, with gains driven primarily by a substantial improvement in recall (\textbf{+7.4\%}) while maintaining, and even slightly improving, precision (\textbf{+1.4\%}). This precision-recall behavior indicates that AVIATOR's augmentation introduces less label noise and achieves broader coverage of vulnerable patterns than VulScribeR. In particular, AVIATOR provides detectors with more learnable and discriminative vulnerable signals, enabling them to identify a wider set of true vulnerabilities (reflected by higher recall) without incurring additional false alarms (as evidenced by stable precision). Given VulScribeR's reported injection success rate of approximately 82\%, the observed precision-recall profile is consistent with AVIATOR generating higher-fidelity vulnerable samples on average, with an estimated success rate of 93\%, thereby strengthening downstream generalization across detector families.

Overall, these findings suggest that the data generated by AVIATOR is of consistently higher quality than existing augmentation baselines, as reflected by its stable and robust downstream performance across models and evaluation settings.

\begin{figure*}[t!]
\centering
\begin{tabular}{cc}
\includegraphics[width=0.49\textwidth]{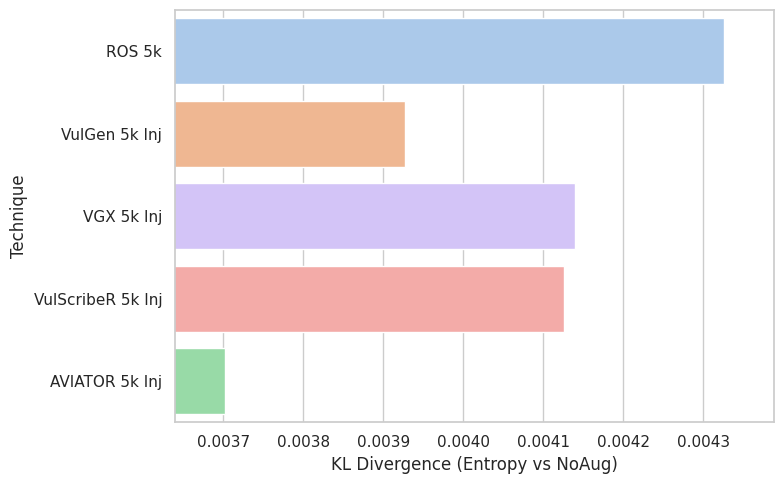} &
\includegraphics[width=0.49\textwidth]{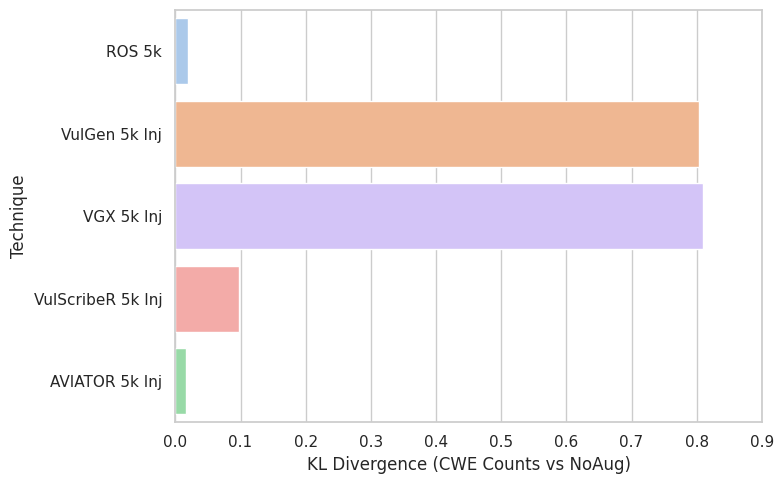} \\
(a) CWE Entropy Distribution Shift to the & (b) CWE Sample Count Distribution\\ Original Dataset (x-axis truncated).  &  Shift to the Original Dataset.
\end{tabular}
\caption{Distributional analysis of augmented datasets. KL divergence measures the deviation from the original, natural distribution of vulnerabilities (lower is better). AVIATOR introduces the least distributional distortion.}
\label{fig:kl_divergence_charts}
\end{figure*}

\paragraph*{Distributional fidelity and realism}
To further assess the realism of AVIATOR-generated data and explain its downstream gains, we analyzed the distributional properties of the augmented datasets. Figure~\ref{fig:kl_divergence_charts} reports the Kullback-Leibler (KL) divergence between the original training data and the augmented data, using two views: sample counts and entropy distributions across CWE categories. A lower KL divergence indicates that the augmentation strategy preserves the natural shape and diversity distribution of real-world datasets, rather than introducing artificial distortion.

AVIATOR achieves the lowest KL divergence under both metrics, which confirms that it remains closest to the original distribution. VulGen and VGX introduce notably larger shifts, consistent with augmentation that concentrates on a narrower set of injection patterns. VulScribeR reduces this distortion compared to pattern-mining baselines but still deviates more than AVIATOR. Overall, these distributional results align with the precision-recall trends in Table~\ref{tab:avg_improvement}: lower distortion corresponds to more realistic, learnable injected vulnerabilities that improve generalization without inducing excessive false alarms.

\paragraph*{Operational superiority: quality, cost, and efficiency}
Beyond downstream metrics, AVIATOR demonstrates significant operational advantages over the state-of-the-art \textit{VulScribeR} in terms of data quality and cost-efficiency.

\begin{itemize}
    \item \textbf{High Syntactic Validity:} We utilized the fuzzy C parser \textit{Joern} \cite{joern_yamaguchi2014} to verify the validity of the augmented samples, following the methodology in \cite{VulScribeR_2025}. While VulScribeR reports a filtering rate (samples rejected due to syntax errors) of \textbf{2\%--13\%}, AVIATOR achieves a significantly lower rejection rate: \textbf{1.13\% for Devign} and \textbf{1.98\% for PrimeVul}. This confirms that AVIATOR's agentic workflow produces more usable training samples with less formatting and parsing noise.
    
    \item \textbf{Cost Efficiency:} AVIATOR achieves these results at a fraction of the cost. While VulScribeR relies on GPT-3.5 Turbo (\$1.88 per 1k samples), AVIATOR's optimized open-weight workflow costs only \textbf{\$0.44 per 1k samples}. This represents a \textbf{4.3$\times$ cost reduction}, making it a far more scalable solution for dataset generation.
\end{itemize}

\begin{center}
\begin{tcolorbox}[myrqstyle]
\textbf{Answer to RQ5:} AVIATOR delivers the strongest \althree{average} augmentation effect in our downstream evaluation\althree{, with the best $F_1$ score in 6 of 8 model--dataset configurations}. It improves average $F_1$ score by \textbf{+21.8\%} over no augmentation and by \textbf{+12.5\%} over ROS, and it outperforms all injection baselines \althree{on average}, including the prior state-of-the-art LLM-based method VulScribeR (\textbf{+2.9\%} average $F_1$). The simultaneous gains in precision and recall, together with the smallest distributional distortion from the original real-world dataset (lowest KL divergence), indicate that AVIATOR generates more realistic and lower-noise vulnerable samples that generalize better across downstream detectors. AVIATOR achieves this quality efficiently, with a $\mathbf{<2\%}$ syntax rejection rate and a \textbf{4.3$\times$} lower cost than VulScribeR. \al{Its value therefore lies in this combination of label fidelity, distributional realism, and efficiency rather than in the $F_1$ margin alone.}
\end{tcolorbox}
\end{center}

\subsection{\altwo{Toward Other Programming Languages}}
\label{subsec:other_languages}

\al{We evaluated AVIATOR on C/C++ in accordance with prior work \cite{vulgen_nong2023, vgx_nong2024, VulScribeR_2025}. The workflow itself, however, was built to be generic: its behavior is driven by model reasoning and self-iterating logic that remains largely independent of the target programming language. Porting it therefore follows a fixed recipe. Within the workflow, the only language-coupled component is the Lightweight Static Analyzer, which we designed to be swapped (Section~\ref{subsec:valid_module}); the agents, their prompts, and the control flow remain unchanged. The method further requires a base LLM fluent in the target language and a set of benign functions to serve as injection targets, both readily available for widely used languages. Fine-tuning the injection agent is optional, as the workflow already performs well without it (Section~\ref{sec:results}). The remaining effort lies in the evaluation scaffolding, namely the oracle and the benchmark used to measure injection success, which must be re-instantiated for each language.}

\altwo{We instantiated this recipe for Java. We replaced Cppcheck with \textit{PMD}\footnote{\url{https://pmd.github.io/}} and \textit{SpotBugs}\footnote{\url{https://spotbugs.github.io/}} as the Lightweight Static Analyzer, and ESBMC with \textit{JBMC} \cite{jbmc_cordeiro2018} as the verification oracle. JBMC belongs to the same family of SMT-based bounded model checkers as ESBMC, so the same bound and timeout protocol applies. As a benchmark, we used the Juliet Java test suite \cite{juliet_java2017}, randomly selecting 100 benign/vulnerable pairs. All agents, prompts, and control flow were kept the same, and we used the base model without fine-tuning.}

\altwo{On this benchmark, AVIATOR reaches an average injection success rate of \textbf{96\%} ($AISR_5$, $\sigma = 2.0$), in line with the 95\% obtained on SARD100 with C/C++ software (Table~\ref{tab:rq1_results}). This experiment signals that that the workflow can transfer well to new languages without modification beyond the static analyzer.}

\section{Discussion}

This section discusses why AVIATOR is effective, its implications for vulnerability dataset construction and downstream detection, and the main limitations of the current study.

\subsection{Why Does AVIATOR Work?}
We identify the main design choices behind AVIATOR's effectiveness: agentic decomposition, targeted model-level adaptation and iterative refinement capabilities, mirroring how human experts inject vulnerabilities in practice. \al{Appendix~\ref{appendix:example} illustrates this on a concrete injection, where AVIATOR jointly weakens two interdependent safety checks rather than making a superficial edit.}

(1) \textbf{Agentic decomposition provides reliable guidance and controllability.} 
AVIATOR decomposes vulnerability injection into 13 specialized agents, each responsible for a well-defined subtask such as code understanding, localization, transformation, diff-based checking, and refinement. This explicit separation enables structured reasoning and targeted self-correction instead of relying on a single monolithic generation step. Empirically, the workflow design drives most of the gains: injection success on FormAI increases from 31\% (single-step baseline) to 91\%, and on SARD100 from 84\% to 95\%. These results indicate that decomposition and tool-augmented verification improve injection fidelity.

(2) \textbf{Targeted fine-tuning strengthens the core transformation step.} 
Instead of retraining the full pipeline, we fine-tune only the Vulnerability Injector with LoRA-based SFT. This lightweight adaptation (under 10 hours on a single A100 GPU) improves FormAI from 85\% to 91\% while reducing variance. The model is trained on PrimeVul, a dataset entirely disjoint from the test sets, showing that this adaptation generalizes well across domains.

(3) \textbf{AVIATOR leverages iterative workflow refinement to improve injection robustness.} Agents operate in feedback loops to enhance injection robustness. AVIATOR explicitly detects and corrects common failure modes in LLM code editing (e.g., cosmetic edits, truncation, or insufficient modifications). For example, the Diff Agent can trigger additional refinement passes when changes are overly disruptive or insufficient. This mechanism improves robustness by avoiding brittle, single-statement edits and contributes to consistent performance even without fine-tuning, provided the base LLM has reasonable language coverage. As a result, AVIATOR can adapt to new vulnerability classes and languages, while maintaining diversity in injected flaws.

(4) \textbf{High-quality injection translates into downstream generalization.}
KL-divergence analysis shows that AVIATOR induces the smallest shift in CWE sample-count and entropy distributions relative to the original training set. In addition to the high injection success rate, this result supports the downstream precision-recall profile: methods that concentrate augmentation on narrow templates distort the training distribution and increase false positives. AVIATOR instead adds signal while tracking the baseline distribution more closely, which plausibly explains its stronger and more stable $F_1$ score gains across detector families.

(5) \textbf{Operational scalability strengthens practical impact.}
AVIATOR maintains a low syntax rejection rate under Joern-based filtering ($<2\%$) while reducing generation cost by 4.3$\times$ compared to VulScribeR. These properties enable large-scale dataset construction and augmentation without accumulating unusable samples or incurring prohibitive cost, and they make AVIATOR suitable for repeated regeneration under consistent quality controls.

\subsection{\altwo{Cost-Performance Analysis}}

\altwo{AVIATOR's 13-agent decomposition increases the injection success rate without a higher reported generation cost. Compared with VulScribeR~\cite{VulScribeR_2025}, it combines higher injection success, lower syntax rejection, smaller distributional shift, and higher average downstream $F_1$, at \$0.44 per 1{,}000 samples on self-hosted open-weight models against the \$1.88 reported for its closed-model pipeline. Compared with VGX, it achieves higher injection success, substantially broader CWE coverage (133 vs.\ 23), higher average downstream $F_1$, and lower distributional distortion; detectors trained on VGX-augmented data perform on average slightly below the unaugmented baseline (Table~\ref{tab:downstream_results}), consistent with more template-driven, less realistic samples. Taken together, these results indicate that the additional workflow structure yields higher-quality data than the other techniques, along several complementary dimensions.}

\subsection{Limitations and Threats to Validity}

\paragraph*{Benchmark Scope and Generalizability}
Our evaluation focuses on C/C++ using three datasets (SARD100, FormAI, and PrimeVul), which may not capture all real-world diversity. To mitigate this, we combined synthetic and real-world benchmarks of varying complexity. \al{Extension to other languages is discussed separately in \ref{subsec:other_languages}; extending AVIATOR to multi-file contexts remains future work.}
\paragraph*{Function-Level Context Limitations}
Our current instantiation operates and validates injections at the function level. For datasets such as PrimeVul, missing project context (e.g., type definitions, globals, external callees) can make some injected samples not determinable under function-local analysis. Accordingly, our manual evaluation assesses whether the intended weakness is present within the function body and marks cases that depend on unavailable context as unknown, as done in prior studies \cite{sateIV_okun2013}. Ongoing work extends AVIATOR toward repository-level injection and validation \cite{docsympo_lbath2026, cyberforge_lbath2026}.
\paragraph*{Manual Evaluation Bias}
The qualitative analysis of PrimeVul is based on \al{100} manually annotated samples, which may introduce subjectivity. We mitigated this by involving three security researchers and resolving disagreements via consensus.
\paragraph*{Metric Stability Under Sampling}
LLM generation is stochastic, which may affect reliability of metrics like injection success rate. To address this, we performed multiple independent runs and reported average scores $AISR_k$ with standard deviations across all experiments and measured Pass@k with additional runs on the best configurations.
\paragraph*{Generalization of Fine-Tuned Agent}
Our supervised fine-tuning was performed solely on PrimeVul, raising a potential risk of overfitting to its code characteristics. However, PrimeVul includes a wide variety of real-world vulnerabilities across multiple domains. Moreover, the consistent performance improvements on structurally distinct benchmarks (SARD100 and FormAI) provided strong evidence of generalization beyond the training distribution.
\paragraph*{Distribution metrics and interpretation}
Entropy- and KL-based analyses capture distributional distortion and code diversity proxies, not semantic equivalence to real vulnerabilities. These metrics support our downstream conclusions, but they do not replace direct correctness validation. We therefore interpret them as complementary evidence rather than as ground-truth realism measures.
\paragraph*{Cost estimates and hardware dependence}
Our cost comparison focuses on generation-time compute under a specific hardware and parallelization setting. Absolute costs vary with deployment, batching, and model choice. The key point is the relative trend: AVIATOR achieves strong quality while enabling lower-cost scaling than large proprietary API-based LLM augmentation pipelines.

\section{Conclusion}
In this paper, we introduced AVIATOR, an agentic AI framework designed to systematically inject realistic and category-specific vulnerabilities into secure codebases. By leveraging a multi-agent pipeline that replicates the reasoning processes of cybersecurity experts, AVIATOR ensures that vulnerability injection is not only precise but also deeply contextually aware. This approach allows for a nuanced understanding and handling of vulnerabilities, closely mimicking the analytical depth and precision of human experts.

Across three widely used benchmarks (\textit{SARD100}, \textit{FormAI}, and \textit{PrimeVul}), AVIATOR achieves \textbf{high \althree{function-level} injection success rate} (\textbf{93\%} across benchmarks), outperforming prior injection techniques and providing stronger guarantees for constructing benchmark datasets with reliable labels. Beyond injection correctness, AVIATOR also delivers the \textbf{strongest \althree{average} downstream benefits} under the standard DLVD augmentation protocol: injecting 5k samples improves average downstream $F_1$ score by \textbf{+21.8\%} over no augmentation and \textbf{+24.6\%} over VGX, while surpassing the prior downstream state of the art \textbf{VulScribeR} by \textbf{+2.9\%} average $F_1$ and \textbf{+7.4\%} recall without degrading precision\althree{, achieving the best $F_1$ score in 6 of 8 model--dataset configurations}. AVIATOR induces the \textbf{lowest distributional distortion} relative to the original training dataset, supporting the conclusion that it generates \textbf{more realistic and lower-noise} vulnerable samples that generalize across detector families. \al{Taken together, these results indicate that AVIATOR generates high-quality vulnerability datasets.}

AVIATOR is also \textbf{broadly applicable}: our current instantiation supports \textbf{133 CWE types}, exceeding the coverage of prior pattern-mining approaches (e.g., VGX with 23 CWEs), highlighting strong diversity potential for dataset construction. Finally, AVIATOR is \textbf{practical at scale}, maintaining a $\mathbf{<2\%}$ syntax rejection rate (Joern-based filtering) and achieving a \textbf{4.3$\times$} lower generation cost than comparable LLM-based injection pipelines.

Future work will extend \textsc{AVIATOR} beyond function-level edits toward multi-function and project-level injection, enabling validation under richer context and supporting a broader class of real-world vulnerability scenarios.

\section*{Declaration of Competing Interest}
The authors declare that they have no known competing financial interests or personal relationships that could have appeared to influence the work reported in this paper.

\section*{Data Availability}
\al{The full code and datasets used in our evaluation are publicly available at: \url{https://github.com/usnistgov/AVIATOR}}

\section*{Declaration of generative AI and AI-assisted technologies in the manuscript preparation process}
During the preparation of this work the author(s) used Gemini and ChatGPT in order to receive suggestions on wording and form of certain paragraphs. After using this tool/service, the author(s) reviewed and edited the content as needed and take(s) full responsibility for the content of the published article.

\section*{CRediT authorship contribution statement}
\textbf{Amine Lbath:} Writing - Original Draft, Writing - review \& editing, Conceptualization, Methodology, Software, Data Curation, Investigation, Visualization. \textbf{Massih-Reza Amini:} Writing - Original Draft, Writing – review \& editing, Conceptualization, Methodology, Supervision. \textbf{Aurelien Delaitre:} Writing – review \& editing, Conceptualization, Methodology, Supervision, Validation. \textbf{Vadim Okun:} Writing – review \& editing, Conceptualization, Methodology, Supervision, Validation.

\bibliographystyle{unsrt} 
\bibliography{document}

\appendix
\subsection{\al{GRPO Objective}}
\label{appendix:grpo}
\al{This appendix gives the full GRPO objective referenced in Section~\ref{subsec:vuln_inj_module}, using the notation introduced there. The CodeBLEU reward $r(o_i)$ of each sampled candidate is normalized into a group-relative advantage}
{\small
\begin{equation}
\label{eq:grpo_advantage}
A_i = \frac{r(o_i) - \mu}{\sigma},\quad
\mu = \frac{1}{G}\sum_{j=1}^{G} r(o_j),\quad
\sigma = \text{std}\left(\{r(o_j)\}_{j=1}^{G}\right).
\end{equation}
}
\al{and the policy $\pi_\theta$ is updated by maximizing the clipped surrogate objective}
{\small
\begin{multline}
\label{eq:grpo_loss}
\mathcal{J}_{\text{GRPO}}(\theta) = \mathbb{E}\left[
    \frac{1}{G}\sum_{i=1}^{G}
    (
        \min(r_i A_i,\, \text{clip}(r_i, 1{-}\epsilon, 1{+}\epsilon) A_i)
      \right. \\ - \beta\,\mathcal{D}_{\text{KL}}(\pi_{\theta}\parallel \pi_{\text{ref}})
    )
\Bigg]
\end{multline}
}
\al{where $r_i = \frac{\pi_{\theta}(o_i \mid q)}{\pi_{\theta_{\text{old}}}(o_i \mid q)}$, $\epsilon$ controls clipping for stable optimization, and $\beta$ regulates the strength of the KL-divergence penalty with respect to a reference policy $\pi_{\text{ref}}$. The expectation is taken over $q \sim P(Q)$ and $\{o_i\}_{i=1}^{G} \sim \pi_{\theta_{\text{old}}}(O \mid q)$, and the KL penalty keeps policy updates close to the pretrained model:}

\begin{equation}
\label{eq:grpo_KL}
\mathcal{D}_{\text{KL}}(\pi_\theta \,\|\, \pi_{\text{ref}}) = \sum_{o_i}\pi_{\theta}(o_i|q)\log\frac{\pi_{\theta}(o_i|q)}{\pi_{\text{ref}}(o_i|q)}.
\end{equation}

\subsection{\al{Example of an Injected Vulnerability}}
\label{appendix:example}

\al{Figure~\ref{fig:injection_example} shows AVIATOR introducing a CWE-170 weakness into \texttt{stat\_add\_node}, a function from SPICE (present in PrimeVul) that stores a node name in the fixed-size buffer \texttt{node->name}. The benign version stays safe through two mechanisms: a length check that rejects oversized names, and a copy via \texttt{g\_strlcpy}, which always null-terminates. AVIATOR removes the check and replaces \texttt{g\_strlcpy} with \texttt{strncpy}, leaving \texttt{node->name} unterminated when the name fills the buffer, so later reads can run past its end.}

\begin{figure*}[!t]
\centering
\begin{lstlisting}[style=diff]
 StatNodeRef stat_add_node(StatNodeRef parent, const char *name, int visible)
 {
     StatNodeRef ref;
     SpiceStatNode *node;

     spice_assert(name && strlen(name) > 0);
-    if (strlen(name) >= sizeof(node->name)) {
-        return INVALID_STAT_REF;
-    }
     pthread_mutex_lock(&reds->stat_lock);
     ref = (parent == INVALID_STAT_REF ? reds->stat->root_index :
                                         reds->stat->nodes[parent].first_child_index);
     while (ref != INVALID_STAT_REF) {
         node = &reds->stat->nodes[ref];
         if (strcmp(name, node->name)) {
             ref = node->next_sibling_index;
         } else {
             pthread_mutex_unlock(&reds->stat_lock);
             return ref;
         }
     }
     if (reds->stat->num_of_nodes >= REDS_MAX_STAT_NODES || reds->stat == NULL) {
         pthread_mutex_unlock(&reds->stat_lock);
         return INVALID_STAT_REF;
     }
     reds->stat->generation++;
     reds->stat->num_of_nodes++;
     for (ref = 0; ref <= REDS_MAX_STAT_NODES; ref++) {
         node = &reds->stat->nodes[ref];
         if (!(node->flags & SPICE_STAT_NODE_FLAG_ENABLED)) {
             break;
         }
     }
     spice_assert(!(node->flags & SPICE_STAT_NODE_FLAG_ENABLED));
     node->value = 0;
     node->flags = SPICE_STAT_NODE_FLAG_ENABLED | (visible ? SPICE_STAT_NODE_FLAG_VISIBLE : 0);
-    g_strlcpy(node->name, name, sizeof(node->name));
+    strncpy(node->name, name, sizeof(node->name));
     insert_stat_node(parent, ref);
     pthread_mutex_unlock(&reds->stat_lock);
     return ref;
 }
\end{lstlisting}
\caption{\al{AVIATOR injection of CWE-170 on \texttt{stat\_add\_node} (SPICE, PrimeVul). Lines marked \texttt{-} are present in the benign function and removed, and \texttt{+} are introduced by AVIATOR.}}
\label{fig:injection_example}
\end{figure*}

\al{The two safeguards are interdependent: removing either alone leaves the function safe. Injecting the vulnerability therefore requires recognizing both and removing them together while leaving the rest of the function intact, reflecting, in this example, an understanding of the code and of the vulnerability itself.}

\begin{figure*}[h]
    \centering
    \includegraphics[width=\linewidth]{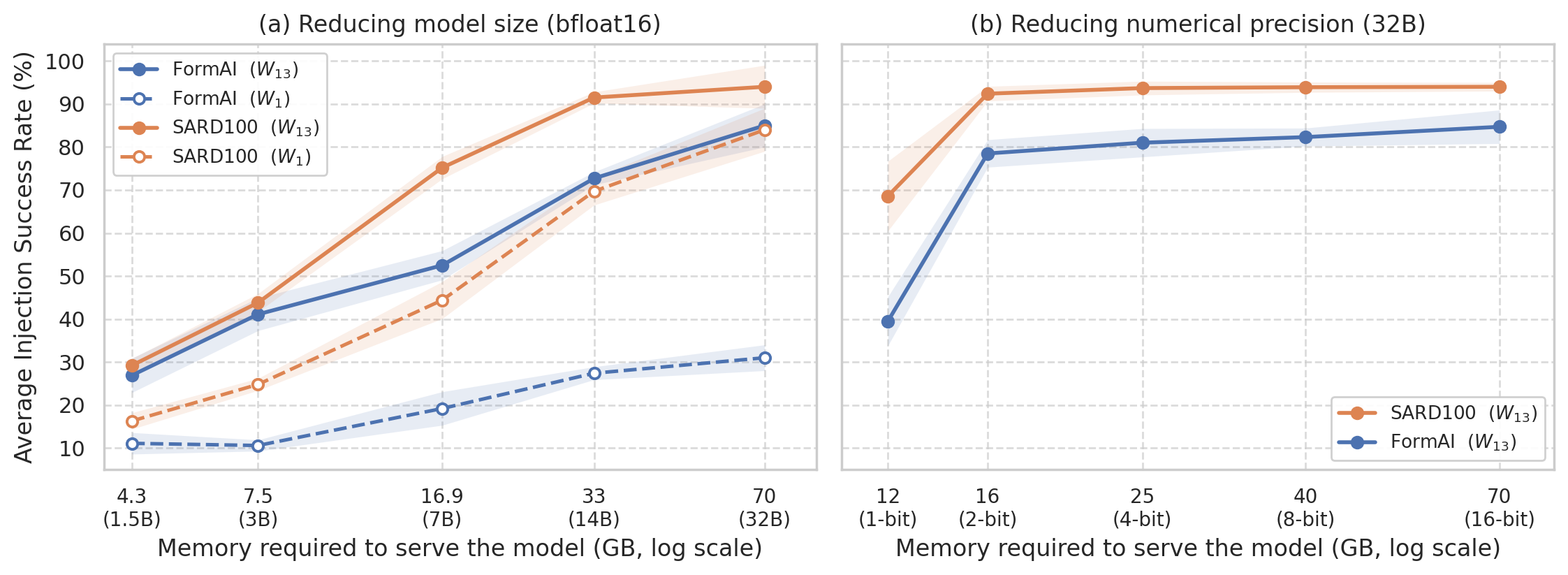}
    \caption{\altwo{Average injection success rate against the memory required to serve the model, for the two ways of meeting a memory budget. (a) Qwen2.5-Coder family at bfloat16, under the single-agent baseline ($W_1$, dashed) and the full workflow ($W_{13}$, solid). (b) The 32B model at five numerical precisions under $W_{13}$, obtained with IQ1\_S, IQ2\_M, AWQ, FP8 and bfloat16 respectively. Model sizes and bit widths are given below the memory values. Bands show the standard deviation over five runs.}}
    \label{fig:scaling}
\end{figure*}

\subsection{\altwo{Injection Success Across Model Sizes and Numerical Precisions}}
\label{appendix:model_sizes}

\altwo{Running AVIATOR on a smaller memory budget can be achieved by using a smaller model or by serving the same model at a lower numerical precision. Figure~\ref{fig:scaling} reports both.}

\altwo{Panel (a) covers four smaller models of the same family, Qwen2.5-Coder at 1.5B, 3B, 7B and 14B parameters, under the single-agent baseline ($W_1$) and the full workflow ($W_{13}$). The workflow improves injection success at every model size. The improvement is largest on the more complex FormAI benchmark, where $W_1$ stays between 11\% and 31\% across the whole range of model sizes while $W_{13}$ reaches 85\%. The benefit of the agentic decomposition is therefore not specific to a single model: workflow design and model capacity are complementary, and injection success increases along both dimensions.}

\altwo{Panel (b) covers the 32B model served at five precisions. Injection success is essentially unchanged down to 4 bits and loses 1.6 points on SARD100 at 2 bits, and it collapses at 1 bit on both benchmarks. At 2 bits, serving the model requires about 16~GB of GPU memory and therefore can run on a single consumer card such as an NVIDIA RTX 3090 or RTX 4090 \footnote{\url{https://www.nvidia.com/en-us/geforce/graphics-cards/}}. At a fixed memory budget, lowering the precision is more effective than reducing the model size: within the same 16~GB, the 32B model at 2 bits reaches 92.4\% on SARD100 and 78.5\% on FormAI, against 75.2\% and 52.5\% for the 7B model at bfloat16, which needs 16.9~GB.}

\begin{figure*}[ht]
    \centering
    \includegraphics[width=0.9\linewidth]{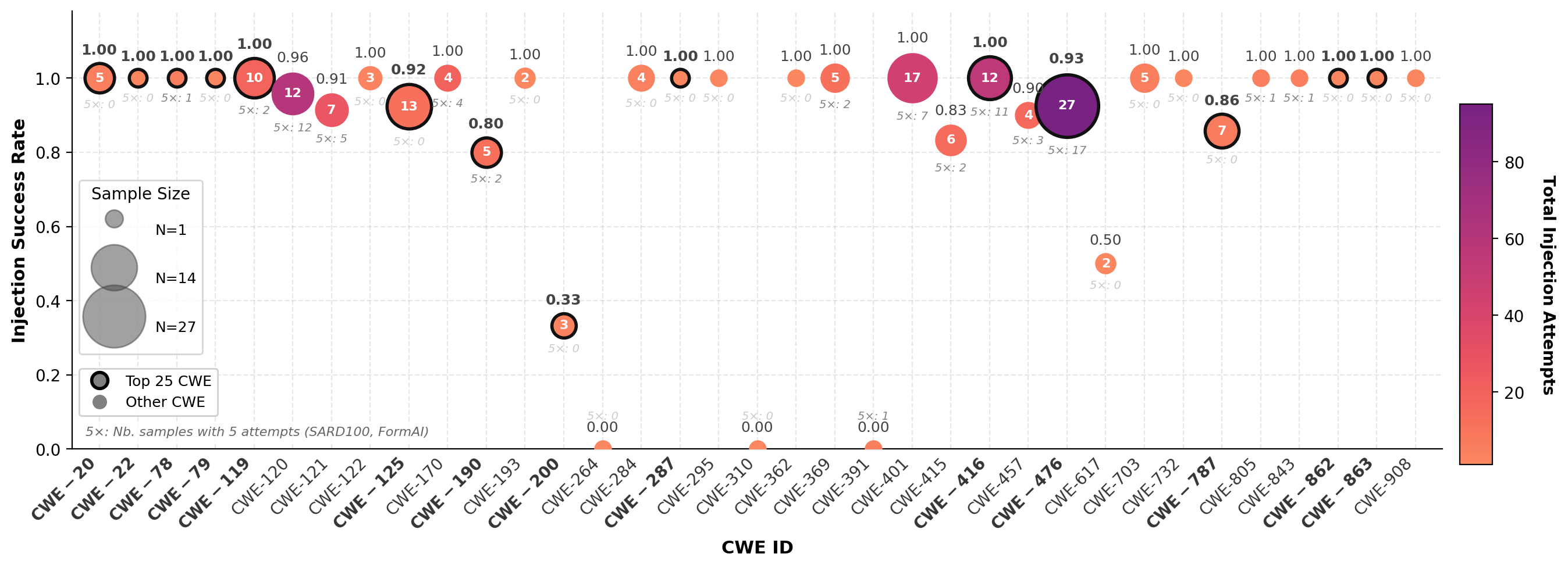}
    \caption{\altwo{Injection success rate by CWE ID within the evaluation datasets. Bubble size indicates the number of unique code samples available for each CWE. The annotation \textit{5×} denotes samples from SARD100 and FormAI evaluated across 5 repeated runs; their reported success rate is the average over the 5 runs. The color gradient reflects the total volume of injection attempts, accounting for all repetitions. CWEs included in the MITRE Top 25 Most Dangerous Software Weaknesses are highlighted with a bold border.}}
    \label{fig:cwe_success_distribution}
\end{figure*}

\subsection{\altwo{Injection Success by CWE}}
\label{appendix:cwe_success}

\altwo{Figure~\ref{fig:cwe_success_distribution} reports the evaluation of Section~\ref{subsec:rq4_variability} at the level of individual CWEs, covering the 35 CWEs present in the evaluation datasets.}

\subsection{\altwo{Replacing a High-Level Ambiguous CWE with a Specific Weakness}}
\label{appendix:deferral}

\altwo{The Confidentiality group in Section~\ref{subsec:rq4_variability} contains four samples labeled CWE-200 or CWE-310. We re-injected the same seed functions using the most specific applicable CWE from AVIATOR's 133-type catalog, selected automatically by the injection model without manual intervention: CWE-252, CWE-209, and CWE-259 for two samples. With these more specific labels, injection success increases from one to three of the four samples.}

\subsection{Prompt Templates Used in AVIATOR}
\label{appendix:prompts}

This appendix documents the exact prompt templates used by the AI agents in the
AVIATOR workflow. Each prompt enforces a strict JSON schema to ensure reliable inter-agent
communication, deterministic parsing, and robust validation across iterative refinement
loops. All prompts are fixed and reused across experiments.

All prompts utilize the same system prompt:
\textit{You are a cybersecurity and code analysis expert with deep knowledge in vulnerability detection, secure coding practices, and static code analysis.}

\subsubsection{Code Understanding Agent}
\label{app:code_understanding_agent}

\begin{tcolorbox}[
  enhanced,
  breakable,
  colback=white,
  colframe=black,
  title=\textbf{Prompt Template: Code Understanding Agent}
]

\textbf{Your task is to analyze the benign function below and try to better understand its goals and its code structure.}

\medskip
\texttt{\{benign\_code\}}

\medskip
\textbf{Return your response as a valid JSON object with exactly the following parameters, maintaining correct types:}

\begin{itemize}
  \item \texttt{"function\_purpose"}: \textit{(str)} Provide a proper description of what the code above is intended to do.
  
  \item \texttt{"input\_descriptions"}: \textit{(list[dict[str, str]])} List descriptions explaining all input arguments and what they are used for.  
  Return a JSON list of dictionaries, for example:  
  \texttt{[{"input name": "A string used for ..."}]}.
  
  \item \texttt{"output\_descriptions"}: \textit{(list[dict[str, str]])} List descriptions explaining all return values.
  
  \item \texttt{"source\_sink\_list"}: \textit{(list[str])} Exhaustive list of pairs of corresponding code sources and sinks.  
  A source is a data-entry point (e.g., user input), and a sink is a function that should not be called with unsanitized, untrusted data.  
  One source can have multiple sinks and vice versa, but list them only by pairs.  
  Example:  
  \texttt{[["name of source 1", "name of sink 1"], ["name of source 2", "name of sink 2"]]}.
  
  \item \texttt{"data\_flows"}: \textit{(list[str])} Analyze and describe all the steps of how data propagates between sources and sinks and throughout the whole code.  
  Return a JSON list, where each line is a textual description of all the steps the data follows between one source and one sink.  
  Do this also throughout the entire code.
\end{itemize}

\medskip
\textbf{Do not repeat yourself.}

\medskip
\textbf{Ensure to write the JSON object in between these markers:}  
\verb|```json| and \verb|```|

\end{tcolorbox}

\subsubsection{Vuln.-Specific Code Understanding Agent}
\label{app:prompt_vuln_specific_understanding}

\begin{tcolorbox}[
  enhanced,
  breakable,
  colback=white,
  colframe=black,
  title=\textbf{Prompt Template: Vuln.-Specific Code Understanding Agent},
  fonttitle=\bfseries,
  coltitle=white,
  colbacktitle=black,
  boxrule=0.8pt,
  arc=2pt,
  left=6pt,right=6pt,top=6pt,bottom=6pt
]
Your task is to identify sanitization and vulnerability mitigation elements that make the code below benign.

\medskip
\textbf{The benign source code to analyze is:}
\begin{verbatim}
{benign_code}
\end{verbatim}

\textbf{The purpose of this code is described as:} \texttt{\{function\_purpose\}} \\
\textbf{Its inputs are:} \texttt{\{input\_descriptions\}} \\
\textbf{Its outputs are:} \texttt{\{output\_descriptions\}}

\medskip
Use the following list of sources / sinks and dataflow as starting point to identify the different ways the vulnerability described in \texttt{vul\_inject\_info} is avoided in this benign source code:
\begin{itemize}
  \item \texttt{"source\_sink\_list"}: \texttt{\{source\_sink\_list\}}
  \item \texttt{"data\_flows"}: \texttt{\{data\_flows\}}
\end{itemize}

\textbf{The vulnerability that is prevented in this code can be described as follows:}
\begin{verbatim}
{vul_inject_info}
\end{verbatim}

Return your response as a valid JSON object with exactly the following parameters, maintaining correct types:
\begin{itemize}
  \item \texttt{"sanitization"}: (\texttt{list[dict[str, str]]}) Identify and analyze all sanitization and vulnerability mitigation elements that make the benign code actually non-vulnerable in terms of the vulnerability described above. Return a JSON list of dictionaries \texttt{[\{"description": "Actual explanation here", "code": "corresponding code here"\}]}
\end{itemize}

\textbf{Important:}
\begin{itemize}
  \item Do not include any additional text or formatting outside the JSON object.
  \item Ensure the JSON strictly adheres to the specified parameter names and types.
  \item Ensure to write the JSON object in between these markers
  \textbf{\texttt{\textasciigrave\textasciigrave\textasciigrave json}}
{\scriptsize
\begin{verbatim}
{
  "sanitization": [
    {"description": "Actual explanation here", 
     "code": "corresponding code here"},
     ...
  ]
}
\end{verbatim}
}
  \textbf{and} \textbf{\texttt{\textasciigrave\textasciigrave\textasciigrave}}
  \item Be sure to close every string you open with \texttt{"}
  \item Do not ever repeat yourself.
\end{itemize}

Now, please generate the JSON assessment.
\end{tcolorbox}

\subsubsection{Vulnerability Injector Agent}

\begin{tcolorbox}[
  enhanced,
  breakable,
  colback=white,
  colframe=black,
  coltitle=white,
  title={Prompt Template: Vulnerability Injector Agent},
  fonttitle=\bfseries,
  colbacktitle=black,
  boxrule=0.8pt,
  arc=2mm,
  left=3mm,right=3mm,top=2mm,bottom=2mm
]
Your task is to modify the source code of non-vulnerable function below to introduce the requested vulnerability.

\medskip
\textbf{The non-vulnerable code that must be modified to introduce changes that will make this code vulnerable is:}
\begin{verbatim}
{benign_code}
\end{verbatim}

\textbf{The purpose of this code is described as:} \{function\_purpose\}

\medskip
\textbf{Its data flows are:} \{data\_flows\}

\medskip
\textbf{The list of sanitization and vulnerability mitigation elements that made the benign code actually non-vulnerable are:}
\begin{verbatim}
{sanitization}
\end{verbatim}

\medskip
\textbf{Think step by step} and leverage the previous analysis of the code to change the code to introduce this vulnerability: \{vul\_inject\_info\}

\medskip
\textbf{As a model you can leverage these examples of how a benign code was modified to introduce the vulnerability \{vul\_inject\_id\}:}
\begin{verbatim}
{retrieved_context}
\end{verbatim}

\medskip
Return your response as a valid JSON object with exactly the following parameters, maintaining correct types:
\begin{itemize}
  \item \texttt{"vulnerable\_code"}: (str) Modified source code with vulnerability introduced. Write your code in between these tags \texttt{<raw> full vulnerable code </raw>}. Rewrite the full code, do not truncate by adding comments such as \texttt{"//... rest of the code..."}. Do not add comments indicating where the vulnerability was introduced.
\end{itemize}

\textbf{Important:}
\begin{itemize}
  \item Do not include any additional text or formatting outside the JSON object.
  \item Ensure the JSON strictly adheres to the specified parameter names and types.
  \item Ensure to write the JSON object in between these markers \verb|```json| and \verb|```|
  \item Ensure to write the code in between these tags \texttt{<raw> full vulnerable code </raw>}
  \item Write the code for the full function with modifications where the original benign code is changed to introduce a new vulnerability.
  \item Output only a valid JSON object. Do not include markdown formatting (no triple backticks or extra quotes). Please ensure that any internal quotes are escaped properly.
  \item Do not ever repeat yourself.
\end{itemize}

Now, please modify the benign function accordingly and output only the JSON object as specified. Be sure to finish the JSON with \verb|```|
\end{tcolorbox}

\subsubsection{Polisher Agent}
\label{app:prompt_polisher_agent}

\begin{tcolorbox}[
  enhanced,
  breakable,
  colback=white,
  colframe=black,
  title={\textbf{Prompt Template: Polisher Agent}},
  fonttitle=\bfseries,
  boxrule=0.8pt,
  arc=2mm,
  left=3mm,right=3mm,top=2mm,bottom=2mm
]
\textbf{You are tasked to verify that the benign code below has been properly modified to introduce the CWE vulnerability \{vul\_inject\_id\}.}

\medskip
\textbf{The initial non-vulnerable source code function was:}
\begin{tcbverbatimwrite}{\jobname_tmp_polisher_benign.txt}
{benign_code}
\end{tcbverbatimwrite}
\verbatiminput{\jobname_tmp_polisher_benign.txt}

\medskip
\textbf{The new version of the code that must have been modified to introduce changes that supposedly make this code vulnerable:}
\begin{tcbverbatimwrite}{\jobname_tmp_polisher_vuln.txt}
{vulnerable_code}
\end{tcbverbatimwrite}
\verbatiminput{\jobname_tmp_polisher_vuln.txt}

\medskip
\textbf{Verify that the code was properly modified to introduce the following vulnerability:} \{vul\_inject\_info\}.

\medskip
\textbf{You must ask yourself the following questions and think step by step as an expert:}
\begin{itemize}
  \item You claimed that introducing the following changes would transform the code to become vulnerable. Does this really introduce the CWE vulnerability \{vul\_inject\_id\} and actually modify the code?
  \item If yes, did you actually made the change? Are you sure?
  \item If no, can you modify the initial non-vulnerable source code function to introduce the requested vulnerability.
  \item Either way, verify that the vulnerability was introduced in a realistic way, as if a human developer had made a mistake.
  \item Make sure the code still makes sense and did not lose its functionality.
  \item Have you truncated the code to only show the modifications? The vulnerable code should show the full code.
\end{itemize}

\textbf{If you need to make some modifications, rewrite the values. Otherwise, if everything is correct, return the values without changes.}

\medskip
\textbf{Return your response as one valid JSON object with exactly the following parameters, maintaining correct types:}
\begin{itemize}
  \item \texttt{"vulnerable\_code"}: (str) Modified source code with vulnerability introduced. Write your code in between these tags \texttt{\detokenize{<raw>}} full vulnerable code \texttt{\detokenize{</raw>}}. Make sure you checked the questions above and the code is properly vulnerable.
  \item \texttt{"injection\_location"}: (str) Location where you have modified the code to introduce the vulnerability (line number, function section and explanation of the location).
  \item \texttt{"injection\_justification"}: (str) Detailed explanation of why you have modified the code to introduce this vulnerability in this specific way, referencing function analysis and retrieved context.
  \item \texttt{"confidence\_score"}: (int) Score between 0 and 10 on how confident you are that your answer is correct and the vulnerable code actually is vulnerable with the proper vulnerability.
  \item \texttt{"confidence\_justification"}: (str) Justification of the confidence score.
\end{itemize}

\medskip
\textbf{Important:}
\begin{itemize}
  \item Do not include any additional text or formatting outside the JSON object.
  \item Ensure the JSON strictly adheres to the specified parameter names and types. Do not write the types.
  \item Ensure to write the JSON object in between these markers \verb|```json| and \verb|```|.
  \item Output only a valid JSON object. Do not include markdown formatting (no triple backticks or extra quotes). Please ensure that any internal quotes are escaped properly.
  \item Do not ever repeat yourself.
\end{itemize}

\textbf{Now, please make the verifications and output only the JSON object as specified.}
\end{tcolorbox}

\subsubsection{Diff Refiner Agent}
\label{app:diff_refiner_prompt}

\begin{tcolorbox}[
  enhanced,
  breakable,
  colback=white,
  colframe=black,
  title={\textbf{Prompt Template: Diff Refiner Agent}},
  fonttitle=\bfseries,
  coltitle=white,
  colbacktitle=black,
  boxrule=0.8pt,
  arc=2mm,
  left=3mm,right=3mm,top=2mm,bottom=2mm
]
You are tasked with verifying that the changes made to the code below actually correspond to the requested vulnerability and that the generated vulnerable code is not missing any parts from the original code that would change its functionality.

\medskip
\textbf{The original benign code is:}
\begin{verbatim}
{benign_code}
\end{verbatim}

\textbf{The vulnerable code that was produced is:}
\begin{verbatim}
{vulnerable_code}
\end{verbatim}

\textbf{The vulnerability to introduce is:} \\
\texttt{{vul\_inject\_id}}: \texttt{{vul\_inject\_info}}

\medskip
\textbf{The purpose of this code is described as:} \texttt{{function\_purpose}} \\
\textbf{Its data flows are:} \texttt{{data\_flows}}

\medskip
\textbf{The list of sanitization and vulnerability mitigation elements that made the benign code non-vulnerable are:}
\begin{verbatim}
{sanitization}
\end{verbatim}

\medskip
\textbf{Your tasks:}
\begin{enumerate}
  \item Analyze the differences between the benign and vulnerable code.
  \item Ensure that the changes made actually introduce the requested vulnerability.
  \item Ensure that the vulnerable code is not missing any parts from the original code that would change its intended functionality.
  \item If the changes are insufficient or incorrect, update the vulnerable code to properly introduce the vulnerability while preserving the original functionality.
\end{enumerate}

\medskip
Return your response as a valid JSON object with exactly the following parameters, maintaining correct types:
\begin{itemize}
  \item \texttt{"diff\_analysis"}: (str) Analysis of whether the code changes correspond to the requested vulnerability and if any functionality is lost.
  \item \texttt{"injection\_location"}: (str) Location where you have modified the code to introduce the vulnerability (line number, function section and explanation of the location).
  \item \texttt{"injection\_justification"}: (str) Detailed explanation of why you have modified the code to introduce this vulnerability in this specific way, referencing function analysis and retrieved context.
  \item \texttt{"confidence\_score"}: (int) Score between 0 and 10 on how confident you are that your answer is correct and the vulnerable code actually is vulnerable with the proper vulnerability.
  \item \texttt{"confidence\_justification"}: (str) Justification of the confidence score.
\end{itemize}

\medskip
\textbf{Important:}
\begin{itemize}
  \item Do not include any additional text or formatting outside the JSON object.
  \item Ensure the JSON strictly adheres to the specified parameter names and types.
  \item Ensure to write the JSON object in between these markers \texttt{\textasciigrave\textasciigrave\textasciigrave json} and \texttt{\textasciigrave\textasciigrave\textasciigrave}.
  \item Ensure to write the code in between these tags \texttt{<raw> full vulnerable code </raw>}.
  \item Output only a valid JSON object. Do not include markdown formatting (no triple backticks or extra quotes). Please ensure that any internal quotes are escaped properly.
  \item Do not ever repeat yourself.
\end{itemize}

\medskip
Now, please analyze and, if needed, update the vulnerable code and output only the JSON object as specified. Be sure to finish the JSON with \texttt{\textasciigrave\textasciigrave\textasciigrave}.
\end{tcolorbox}

\subsubsection{Diff Analyzer Agent}
\begin{tcolorbox}[enhanced, breakable, colback=white,colframe=black,title={\textbf{Prompt Template: Diff Analyzer}}]
You need to critically analyze the code changes made to introduce a vulnerability and verify that they properly implement the intended vulnerability type.

\textbf{The original benign code was:}
\begin{verbatim}
{benign_code}
\end{verbatim}

\textbf{The vulnerable code that was produced is:}
\begin{verbatim}
{vulnerable_code}
\end{verbatim}

\textbf{The difference between the benign and vulnerable code is shown below with [ADDED] and [REMOVED] markers:}
\begin{verbatim}
{code_diff}
\end{verbatim}

\textbf{The vulnerability that should have been introduced is:}
\begin{verbatim}
{vul_inject_id}: {vul_inject_info}
\end{verbatim}

\textbf{You are claiming to have made the following modification:}
\begin{itemize}
  \item Location: \texttt{\{injection\_location\}}
  \item Justification: \texttt{\{injection\_justification\}}
\end{itemize}

\textbf{Your task is to critically analyze whether:}
\begin{enumerate}
  \item The modifications actually match what was claimed in the injection\_location
  \item The introduced changes actually correspond to the vulnerability described in vul\_inject\_info
  \item The vulnerability has been properly implemented, not just superficially simulated
  \item The code changes are substantive and not merely cosmetic
\end{enumerate}

\textbf{Return your response as a valid JSON object with exactly the following parameters, maintaining correct types:}
\begin{itemize}
  \item \texttt{analysis\_result}: (str) Your detailed critical analysis explaining whether the code modification properly introduces the intended vulnerability. Include specific references to the code changes and vulnerability requirements.
  \item \texttt{modification\_valid}: (bool) A boolean value indicating whether the modification properly introduces the intended vulnerability (true) or if it fails to do so (false).
\end{itemize}

\textbf{Important:}
\begin{itemize}
  \item Do not include any additional text or formatting outside the JSON object.
  \item Ensure the JSON strictly adheres to the specified parameter names and types.
  \item Ensure to write the JSON object in between these markers \texttt{\detokenize{```json}} and \texttt{\detokenize{```}}
  \item Be critical and thorough in your assessment - the goal is to verify that the vulnerability was properly introduced.
  \item Do not ever repeat yourself.
\end{itemize}

Now, please analyze the code changes and provide your assessment as specified.
\end{tcolorbox}

\subsubsection{Vulnerability Verifier Agent}
\label{app:prompt_vulnerability_verifier_agent}

\begin{tcolorbox}[
  enhanced,
  breakable,
  colback=white,
  colframe=black,
  title={\textbf{Prompt Template: Vulnerability Verifier Agent}},
  fonttitle=\bfseries,
  boxrule=0.8pt,
  arc=2mm,
  left=3mm,right=3mm,top=2mm,bottom=2mm
]

\textbf{Role.} You are a comprehensive vulnerability verification system with advanced knowledge of common programming flaws, security issues, and code vulnerabilities.

\medskip
\textbf{Goal.} Perform a \textbf{final validation} of whether the intended vulnerability has been properly introduced in the code, using all available evidence.

\medskip
\textbf{Inputs.}
\begin{itemize}
  \item \textbf{Original benign code:} \texttt{\{benign\_code\}}
  \item \textbf{Produced vulnerable code:} \texttt{\{vulnerable\_code\}}
  \item \textbf{Intended vulnerability:} \texttt{\{vul\_inject\_id\}}: \texttt{\{vul\_inject\_info\}}
  \item \textbf{Prior assessments:}
  \begin{itemize}
    \item Critical analysis result: \texttt{\{analysis\_result\}}
    \item Modification validity: \texttt{\{modification\_valid\}}
  \end{itemize}
  \item \textbf{Static analysis signals:}
  \begin{itemize}
    \item CWE IDs detected: \texttt{\{cwe\_ids\}}
    \item Static analysis errors: \texttt{\{static\_analysis\_errors\}}
  \end{itemize}
  \item \textbf{Diff evidence:} \texttt{\{code\_diff\}} (with \texttt{[ADDED]} / \texttt{[REMOVED]} markers)
  \item \textbf{Claimed modification:}
  \begin{itemize}
    \item Location: \texttt{\{injection\_location\}}
    \item Justification: \texttt{\{injection\_justification\}}
  \end{itemize}
\end{itemize}

\medskip
\textbf{Your tasks (must follow all).}
\begin{enumerate}
  \item Evaluate whether the \textbf{CWE detection} matches the intended vulnerability type.
  \item Cross-reference the \textbf{code diff} with the \textbf{static analysis} findings.
  \item Determine whether the vulnerable code exhibits the weakness described in \texttt{\{vul\_inject\_info\}}.
  \item Verify that the modification is \textbf{realistic} and \textbf{meaningful} (not cosmetic).
  \item Confirm that the intended vulnerable pattern has been implemented \textbf{correctly}.
\end{enumerate}

\medskip
\textbf{Output format.} Return \textbf{one valid JSON object} with exactly the following parameters (maintaining correct types):
\begin{itemize}
  \item \texttt{"is\_correctly\_vulnerable"}: (bool) Whether the code correctly implements the intended vulnerability.
  \item \texttt{"verification\_explanation"}: (str) Detailed explanation with evidence from static analysis, diff, and prior assessments.
\end{itemize}

\medskip
\textbf{Important:}
\begin{itemize}
  \item Do \textbf{not} include any additional text or formatting outside the JSON object.
  \item Ensure the JSON strictly adheres to the specified parameter names and types.
  \item Ensure to write the JSON object in between these markers \texttt{\detokenize{```json}} and \texttt{\detokenize{```}}.
  \item Be \textbf{comprehensive and thorough} in your final assessment.
  \item Do not ever repeat yourself.
\end{itemize}

\medskip
Now, please provide your final verification of the vulnerability introduction as specified.

\end{tcolorbox}

\vfill

\subsection{\altwo{Supported CWE Types}}
\label{appendix:supported_cwes}

\begin{table*}[t]
    \centering
    \caption{\altwo{The 133 CWE types supported by the current AVIATOR instantiation.}}
    \label{tab:supported_cwes}
    \footnotesize\setlength{\tabcolsep}{0pt}\renewcommand{\arraystretch}{1.0}
    \begin{tabular}{@{}r@{\hspace{0.6em}}l@{\hspace{3em}}r@{\hspace{0.6em}}l@{}}
        \toprule
        \textbf{CWE} & \textbf{Name} & \textbf{CWE} & \textbf{Name} \\
        \midrule
        16 & Configuration & 362 & Race Condition \\
        19 & Data Handling Issues & 367 & Time-of-Check Time-of-Use (TOCTOU) Race Condition \\
        20 & Improper Input Validation & 369 & Division by Zero \\
        22 & Path Traversal & 388 & Insufficiently Protected Credentials \\
        59 & Link Following & 391 & Unchecked Error Condition \\
        61 & UNIX Symbolic Link (Symlink) Following & 399 & Resource Management Errors \\
        74 & Injection & 400 & Uncontrolled Resource Consumption \\
        77 & Command Injection & 401 & Improper Release of Memory Before Removing Last Reference \\
        78 & OS Command Injection & 404 & Improper Resource Shutdown or Release \\
        79 & Cross-site Scripting & 412 & Unrestricted Externally Accessible Lock \\
        80 & Basic Cross-Site Scripting & 415 & Double Free \\
        88 & Argument Injection or Modification & 416 & Use After Free \\
        89 & SQL Injection & 417 & Unsafe Practices in a Concurrent Environment \\
        93 & OS Command Injection & 426 & Untrusted Search Path \\
        94 & Code Injection & 444 & Inaccessible Code \\
        99 & Resource Injection & 457 & Use of Uninitialized Variable \\
        113 & HTTP Response Splitting & 466 & Return of Pointer Value Outside of Expected Range \\
        116 & Improper Encoding or Escaping of Output & 468 & Incorrect Pointer Scaling \\
        119 & Buffer Overflow & 476 & NULL Pointer Dereference \\
        120 & Buffer Overflow & 489 & Active Debug Code \\
        121 & Stack-based Buffer Overflow & 502 & Deserialization of Untrusted Data \\
        122 & Heap-based Buffer Overflow & 522 & Insufficiently Protected Credentials \\
        125 & Out-of-Bounds Read & 532 & Insertion of Sensitive Information into a Log File \\
        129 & Improper Validation of Array Index & 552 & Files or Directories Accessible to External Parties \\
        131 & Incorrect Calculation of Buffer Size & 601 & Open Redirect \\
        134 & Uncontrolled Format String & 611 & XML External Entity \\
        170 & Improper Null Termination & 613 & Insufficient Session Expiration \\
        172 & Encoding Error & 617 & Reachable Assertion \\
        189 & Numeric Errors & 639 & Authorization Bypass Through User-Controlled Key \\
        190 & Integer Overflow or Wraparound & 664 & Improper Control of a Resource Through its Lifetime \\
        193 & Off-By-One Error & 665 & Improper Initialization \\
        197 & Numeric Truncation Error & 667 & Improper Locking \\
        200 & Exposure of Sensitive Information & 668 & Insufficient Detection of Unusual or Exceptional Conditions \\
        203 & Observable Discrepancy & 670 & Access to Critical Private Variable via Public Method \\
        209 & Information Exposure Through an Error Message & 672 & Access to Critical Private Variable via Public Method \\
        212 & Path Traversal & 674 & Uncontrolled Recursion \\
        241 & Use of Implicit Trust in a Function & 681 & Incorrect Conversion between Numeric Types \\
        252 & Unchecked Return Value & 682 & Incorrect Calculation \\
        254 & Unchecked Return Value & 690 & Unchecked Return Value to NULL Pointer Dereference \\
        255 & Improper Handling of Excessive Authentication Attempts & 693 & Protection Mechanism Failure \\
        259 & Hardcoded Password & 697 & Incorrect Comparison \\
        264 & Permissions, Privileges, and Access Control & 703 & Improper Check or Handling of Exceptional Conditions \\
        269 & Improper Privilege Management & 704 & Incorrect Type Conversion or Cast \\
        273 & Improper Check for Dropped Privileges & 707 & Improper Neutralization \\
        276 & Incorrect Default Permissions & 732 & Incorrect Permission Assignment for Critical Resource \\
        281 & Incorrect Authentication & 754 & Improper Check for Unusual or Exceptional Conditions \\
        284 & Improper Access Control & 755 & Improper Handling of Exceptional Conditions \\
        285 & Improper Authorization & 763 & Release of Invalid Pointer or Resource \\
        287 & Improper Authentication & 770 & Allocation of Resources Without Limits or Throttling \\
        288 & Authentication Bypass by Improper Validation & 772 & Access to Critical Private Variable via Public Method \\
        290 & Authentication Bypass Using Reverse Tabnabbing & 786 & Access of Memory Location Before Start of Buffer \\
        294 & Connection Pooling Without Validation & 787 & Out-of-bounds Write \\
        295 & Improper Certificate Validation & 788 & Access of Memory Location After End of Buffer \\
        307 & Improper Restriction of Excessive Authentication Attempts & 798 & Use of Hard-coded Credentials \\
        310 & Cryptographic Issues & 805 & Buffer Access with Incorrect Length Value \\
        311 & Improper Access Control & 823 & Untrusted Pointer Dereference \\
        320 & Key Management Errors & 824 & Access to a Restricted Filename or Directory \\
        326 & Inadequate Encryption Strength & 834 & Excessive Computation \\
        327 & Use of a Broken or Risky Cryptographic Algorithm & 835 & Infinite Loop \\
        331 & Insufficient Entropy & 843 & Access to a Memory Location Through a Dereferenced Pointer \\
        345 & Insufficient Verification of Data Authenticity & 862 & Missing Authorization \\
        346 & Origin Validation Error & 863 & Incorrect Authorization \\
        347 & Improper Verification of Cryptographic Signature & 908 & Untrusted Pointer Dereference \\
        349 & Command Injection & 909 & Uncontrolled Resource Consumption \\
        352 & Cross-Site Request Forgery & 918 & Server-Side Request Forgery \\
        354 & Path Traversal & 1021 & Improper Restriction of Rendered UI Layers or Frames \\
        361 & Time-of-Check Time-of-Use (TOCTOU) Race Condition &  &  \\
        \bottomrule
    \end{tabular}
\end{table*}

\subsection{\altwo{SATE V Weakness Groups}}
\label{appendix:sate5_groups}

\begin{table*}[h]
    \centering
    \caption{\altwo{SATE~V weakness groups~\cite{sate5_delaitre2018} and the supported CWE types of Table~\ref{tab:supported_cwes} belonging to each group.}}
    \label{tab:sate5_groups}
    \scriptsize\setlength{\tabcolsep}{0pt}\renewcommand{\arraystretch}{1.15}
    \begin{tabular}{@{}l@{\hspace{0.6em}}r@{\hspace{0.6em}}l@{\hspace{1.6em}}l@{\hspace{0.6em}}r@{\hspace{0.6em}}l@{}}
        \toprule
        \textbf{Weakness Group} & \textbf{\#} & \textbf{Supported CWEs} & \textbf{Weakness Group} & \textbf{\#} & \textbf{Supported CWEs} \\
        \midrule
        Access control & 7 & 264, 284, 285, 552, 639, 668, 732 & Expired memory & 2 & 415, 416 \\
        Ante buffer operation & 8 & 119, 125, 129, 466, 786, 787, 805, 823 & Initialization & 5 & 457, 665, 824, 908, 909 \\
        Authentication & 2 & 346, 613 & Input validation & 12 & 20, 74, 77, 78, 88, 89, 94, 99, 116, 134, 611, 707 \\
        Calculation & 5 & 131, 193, 369, 468, 682 & Invalid pointer & 6 & 466, 476, 690, 763, 823, 824 \\
        Cleanup & 1 & 404 & Loop and recursion & 2 & 674, 835 \\
        Code quality & 1 & 489 & Memory leak & 1 & 401 \\
        Comparison & 1 & 697 & Memory release & 1 & 763 \\
        Concurrency & 3 & 362, 367, 667 & Numeric errors & 5 & 189, 190, 197, 681, 682 \\
        Confidentiality & 8 & 200, 209, 310, 311, 326, 327, 552, 668 & Path-related & 4 & 20, 22, 59, 426 \\
        Control flow & 2 & 617, 670 & Post buffer operation & 12 & 119, 120, 121, 122, 125, 129, 170, 466, 787, 788, 805, 823 \\
        Credentials management & 4 & 255, 259, 613, 798 & Privileges & 2 & 269, 273 \\
        Data structure & 2 & 170, 707 & Resource management & 10 & 99, 399, 400, 404, 664, 672, 770, 772, 908, 909 \\
        Denial of Service & 5 & 400, 401, 404, 674, 770 & Return value & 3 & 252, 273, 690 \\
        Design and implementation & 1 & 693 & Strings & 1 & 134 \\
        Dynamic code & 1 & 94 & Type-related & 3 & 681, 704, 843 \\
        Encapsulation & 1 & 489 & Unhandled errors & 5 & 273, 391, 703, 754, 755 \\
        Environment induced & 2 & 16, 426 & Web & 8 & 20, 79, 80, 113, 346, 352, 601, 611 \\
        Error condition & 2 & 388, 617 &  & &  \\
        \bottomrule
    \end{tabular}
\end{table*}

\end{document}